\documentclass[a4paper,10pt]{article}

\usepackage[utf8]{inputenc}
\usepackage[english]{babel}
\usepackage{amsmath,amsfonts,amssymb}
\usepackage[pdftex]{color,graphicx,hyperref}
\usepackage{subfig}
\usepackage{booktabs}
\DeclareMathOperator{\diag}{diag}

\begin{document}

\centerline{\Huge Criterions for locally dense subgraphs}
\vskip0.7cm
\centerline{\huge Gergely Tibély}
\vskip0.3cm
\centerline{\normalsize Institute of Physics and HAS-BME Cond.~Mat.~Group,}
\centerline{\normalsize Budapest University of Technology and Economics,}
\centerline{\normalsize Budapest, Budafoki str. 8., H-1111}
\vskip2.5cm

\begin{abstract}
Community detection is one of the most investigated problems in the field of complex networks. Although several methods were proposed, there is still no precise definition of communities. As a step towards a definition, I highlight two necessary properties of communities, separation and internal cohesion, the latter being a new concept. I propose a local method of community detection based on two-dimensional local optimization, which I tested on common benchmarks and on the word association database.
\end{abstract}

\section{Introduction}

In the last decade interdisciplinary research on complex networks resulted in spectacular development \cite{AlbertBarab}-\cite{DorMend}. It has become clear that networks constructed from diverse complex systems show remarkably similar features. Several aspects were investigated, like clustering \cite{WattsStrogatz}, the degree distribution\cite{degreedistr}, diameter \cite{diam}, \cite{ultrasmall}, spreading processes \cite{spreading}, diffusion \cite{diff}, synchronization \cite{sync}, critical phenomena \cite{crit_phenom} and game theoretical models on complex networks \cite{gametheory}. 

One of the most actively researched questions about complex networks is the one of community detection \cite{SF_review}. Community detection aims at finding dense groups in graphs, like circles of friends in social networks, web pages about the same topic, or substances appearing in the same pathway in metabolic reaction networks. Perhaps the strongest motivation behind the research is that dense groups in the topology are expected to correspond to functions performed by the network, such that one can infer from pure topology to function. While the concept of communities seems intuitively plausible, attempts for an algorithmically useful definition have not been successful yet. The global characterization by modularity \cite{Q} or by random walks \cite{MAS, Infomap}, the local ``weak`` and ``strong`` definitions \cite{Radicchi}, the clique percolation approach \cite{CPM}, or the multiresolution methods \cite{RB,AFG,KSKK} have all increased out understanding of this complex problem but the proliferation of methods of community detection just indicates the difficulty of this issue \cite{SF_review}.

Unfortunately, any precise definition of communities is still lacking, giving rise to innumerable methods using different definitions. Lack of a definition also makes problematic the testing of methods; although there is progress in this issue \cite{SFbenchmark1}, \cite{SFbenchmark2}. Difficulty of the problem is increased by more subtle factors: very often communities occur on a broad scale, they can be ordered in a hierarchical manner, and they may overlap, which make their identification even harder.

After being the subject of active research for several years, it is getting clear that the following stages appear during community detection:
\begin{itemize}
 \item[1] defining the term ``community''; 
 \item[2] finding the objects corresponding to the definition;
 \item[3] determining the significance of the found communities.
\end{itemize}
Although from the theoretical perspective stage 1 is clearly a key issue, it is far from being settled. Several different propositions exists, which are evaluated mostly according to their results on a few benchmarks. This is the stage to be improved in the first place in this paper. Stage 2 is a technical issue, often consisting of some combinatorial optimization method. Its choice is usually a result of a trade-off between speed and quality. Stage 3 should give information about how surprising is the existence of a found community in the actual graph, given some characteristics of the graph like edge density or degree distribution. Although this issue also got some attention \cite{Q_0value}-\cite{stat_signif_Gfeller}, it just began to get widespread application \cite{OSLOM}.

The rest of the paper will focus on the question of definition, so a few remarks about stage 3 are made here. Most community detection methods give no information about the significance of their output, thus forcing the investigator to assume that all results are (equally) significant. This way, the community detection stages 2 and 3 are combined into a single decision whether a particular subgraph is a good enough community or not -- effectively pruning the significance test in practice. The other end of the spectrum, represented by \cite{OSLOM}, builds the definition of communities on statistical significance, which is clearly an improvement. However, it should be noted that the fitness and statistical significance of a subgraph as a community are not synonyms. Statistical significance tells us how surprising a subgraph is, while fitness talks about how close is it to the ideal community. Therefore, the two quantities are complementary and both belong to the description of a community.

\section{Local criteria for communities} \label{s:def}
A fundamental problem of community detection is to define the term ``community''. There are different approaches to this question. One is the algorithmic approach, giving a computational procedure for finding clusters. This naturally incorporates a mathematically precise definition, although different algorithms usually result in diverse definitions, and there is no theoretical framework currently to help their differentiation. Another possibility is to present a general concept, on which a precise definition can be based. In this paper, the latter approach is taken, although an algorithmic realization is also presented.

No definition of communities which is both precise and generally accepted has appeared yet. Currently the description of communities exhausts in the phrase ``nodes having more edges among themselves than to the rest of the graph'' (or equivalent forms). It can be translated roughly to ``statistically significant locally dense subgraphs``. Statistical significance is a quite precise expression, the main problem is with the term ``locally dense``.  For an intuitive picture, it is quite good, but much less than directly transformable to algorithms. Although there is an implicit agreement on that clearly counterintuitive results are not permitted, even a formal list of required properties is missing. However, there are some properties which fit human intuition about locally dense subgraphs\footnote{For brevity, the words ``community'', ``group'' and ``cluster'' will be used from this point as synonyms for ``locally dense subgraph'', omitting the statistical significance from the meaning.}\footnote{It should be noted that the meaning of the term ``community'' can depend on the context; consequently a single definition may not be enough. Here the aim is to describe a particularly intuitive one.}:
\begin{description}
	\item[Separation:] a good community is well-separated from the rest of the graph;
	\item[Cohesion:] a good community is homogeneously well connected inside, i.e. it is hard to separate into two communities.\footnote{The term ``cohesion'' also appeared in \cite{RB}, although there it denotes a quantity with an unrelated concept.}
\end{description}

The separation criterion is quite clear, although there is an important remark: separation should be defined locally, involving only the community under investigation and its immediate neighborhood. Global methods, in which distant regions of the graph can modify a community in order to improve a global fitness value, can produce results violating the human perception about clusters. A famous example is the resolution limit of modularity \cite{reslim, reslim2}.

Although separation is a very intuitive criterion, and famous methods rely on it (see the Appendix), it is not enough in itself. Figs. \ref{fig:2cliques}-\ref{fig:sw_2cliques} illustrate that the distribution of links inside the separated region (the ``shape`` of the subgraph) also matters heavily. Application of current community detection methods to real-world networks confirms that this is a real problem, e.g. tree-like communities can occur, even when the whole network is not tree-like \cite{various_comparison}, \cite{mobilephone_comparison}. 

Both separation and cohesion are required properties of communities. If one neglects cohesion, the result may contain clusters like the one on Fig. \ref{fig:2cliques}. On the other hand, if separation is not taken into account, one may end up chopping a separated subgraph until very cohesive pieces (cliques in the extreme) are obtained, like the triangles on Fig. \ref{fig:sw_2cliques}.

\begin{figure}[!h]
\begin{center}
\subfloat[]{\label{fig:2cliques}\includegraphics*[width=1.5cm]{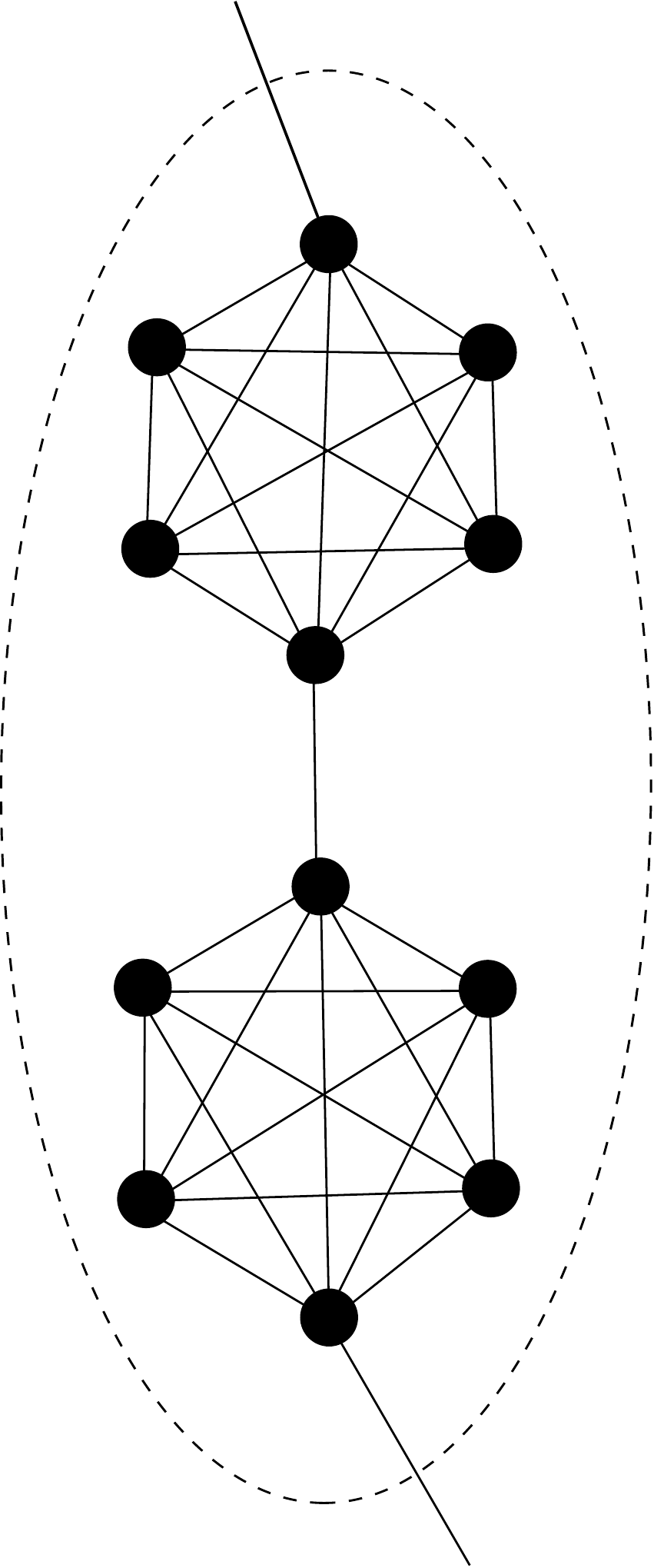}}
\qquad\qquad
\subfloat[]{\label{fig:sw_2cliques}\includegraphics*[width=3cm]{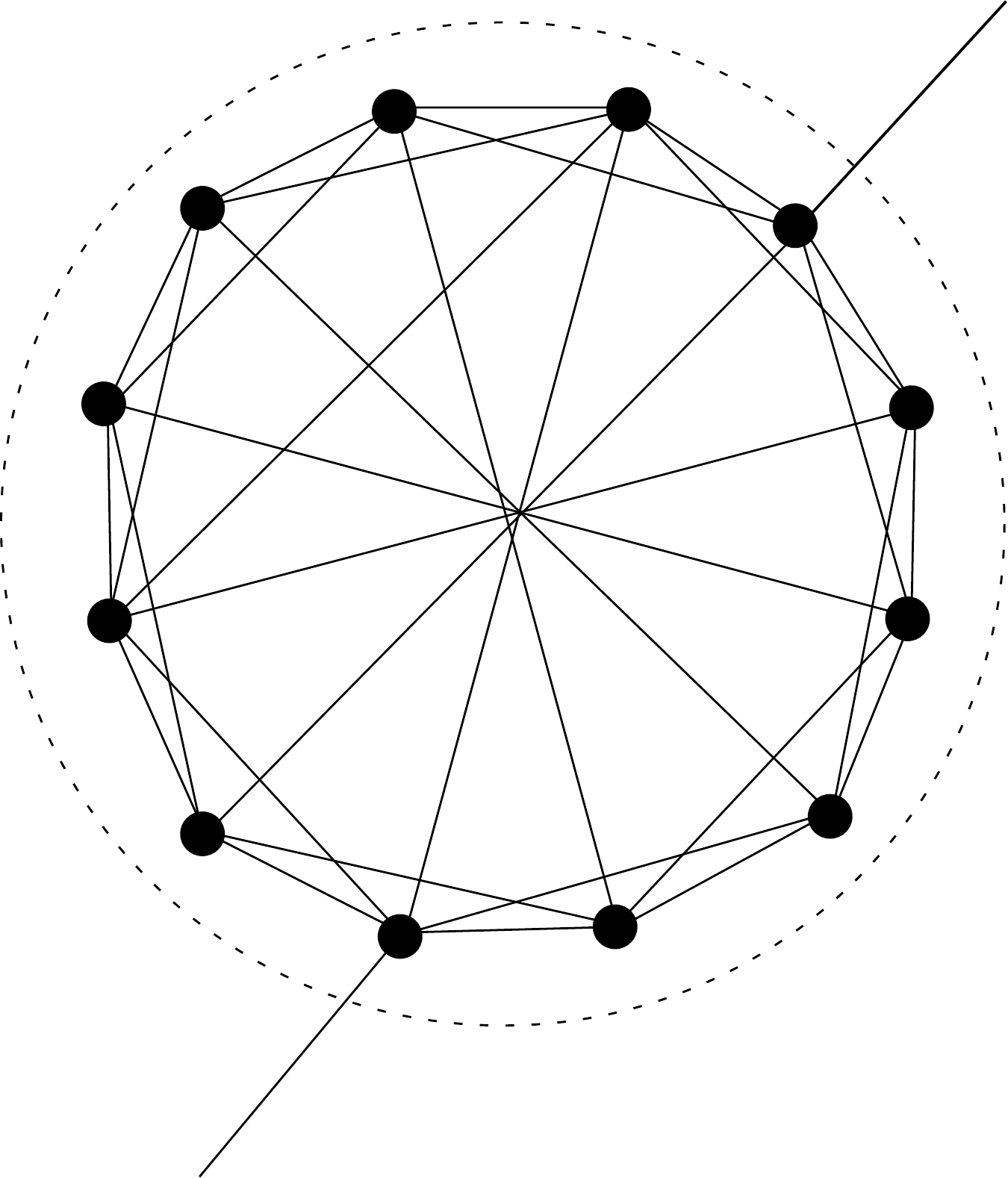}}\\
\end{center}
\caption{Illustration of the importance of subgraph shape. The two subgraphs 
have the same number of nodes and the same degrees, i.e. they differ only in the distribution of links. The figure on the left is much less cohesive than the figure on the right, although just a reorganization was applied to the links.}
\label{fig:counterexample}
\end{figure}

Given the subgraphs on Figs \ref{fig:2cliques} and \ref{fig:sw_2cliques} as proposed communities, most community detection methods' fitness values, to be reviewed in the next section, can not tell the difference between them. This is due to that most methods simply count the internal and/or external edges, which do not tell about the distribution of those edges. The reason why several methods do not fail to assign proper clusters for Fig \ref{fig:2cliques} is that they look for optimal clusters, consequently they compare configurations like Fig \ref{fig:2cliques} in one cluster and in two clusters, and splitting the two cliques into two clusters may improve the partition. But the situation is even worse. In the next Section, we will see that a number of fitness functions are more optimal for a counterintuitive clustering than for the intuitive one (e.g. joining the two cliques on Fig \ref{fig:2cliques}, like modularity for a large enough graph). It should be noted that in such a case, the proper communities might be recovered if the heuristic gets stuck in the proper local optimum, even when that is not the global optimum.

\section{Overview of the existing methods}

Here, existing community detection methods will be reviewed from the point of view of the previous section, i.e. do they conform the criterions of separation and cohesion. As mini-benchmarks, the examples on Fig. \ref{fig:counterexample} or their simple variations will be used (see the Appendix for details on individual methods). The desired output for \ref{fig:2cliques} is two communities consisting of the two cliques, while \ref{fig:sw_2cliques} should be kept in one piece. In both cases, no nodes from the rest of the graph should be included. For methods optimizing a fitness function, the globally optimal solution will be considered, for other methods, the possible solutions. These solutions will be compared to the desired ones, independently for Fig. \ref{fig:2cliques} and \ref{fig:sw_2cliques}. If a method separates the two cliques of Fig. \ref{fig:2cliques}, then it gets a ``+'',  if it puts all nodes of Fig. \ref{fig:sw_2cliques} into one cluster, then it gets another ``+''. If there are multiple equally valid solutions (like for label propagation), all solutions are required to conform the preferred result. \\
For methods optimizing a function, the heuristic realizing the optimization may deviate from the global optimum, presenting worse or even better results (in terms of conformity to separation and cohesion). This will not be investigated, here the focus is on the definition of the communities (following from the choice of the fitness function), not on the practical aspects. Results for methods which can produce a single partition or cover are displayed in Table \ref{t:review}. The large number of published methods makes assembling a complete list nearly impossible. Instead, the emphasis is put on the diversity of the reviewed approaches.

\begin{table}[!h]
 \begin{center}
  \begin{tabular}{ c  c  c  c }
   method & cohesion test & separation test \\
          & (like Fig. \ref{fig:2cliques}) & (like Fig. \ref{fig:sw_2cliques}) \\ 
   \hline
   Lancichinetti et al. \cite{Lancichinetti} & - & + \\
   Labelpropagation \cite{labelprop} & - & - \\
   Infomap \cite{Infomap} & - & + \\
   Clique Percolation \cite{CPM} & - & - \\
   Estrada \& Hatano \cite{Estrada} & - & - \\
   Modularity optimization \cite{Q} & - & + \\
   Donetti \& Mu\~noz \cite{DM} & - & + \\
   Ronhovde \& Nussinov \cite{RN} & - & - \\
   Nepusz et al. \cite{Nepusz} & + & - \\
   Hofman \& Wiggins \cite{HW} & - & - \\
   Hastings \cite{Hastings} & - & - \\
   Newman \& Leicht \cite{NewmanLeicht} & + & - \\ 
   Wang \& Lai \cite{NL++} & + & - \\ 
   Bickel \& Chen \cite{BickelChen} & + & - \\ 
   Karrer \& Newman \cite{Karrer} & + & - \\ 
   Infomod \cite{Infomod} & - & + \\ 
   Radicchi et al. \cite{Radicchi} & + & - \\
   Chauhan et al. \cite{lambda1} & + & - \\ 
   Evans \& Lambiotte \cite{EvansLambiotte} & - & + \\ 
   Ahn et al. \cite{Ahn} & - & - \\ 
   ModuLand \cite{Moduland} & - & - \\
   \hline
  \end{tabular}
  \caption{Cohesion \& separation criterion test results. Tests were done on Fig. \ref{fig:2cliques} and \ref{fig:sw_2cliques} or similar graphs (which are described in the Appendix). + and - are assigned according to whether the fitness function of a method is more optimal for the preferred solution or not. For methods which do not optimize a fitness function, simply the possible solution(s) was (were) analyzed. See the Appendix for details on specific methods.}
  \label{t:review}
 \end{center}
\end{table}

There is a bunch of multiresolution methods, which possess a parameter allowing to tune the cluster sizes from $1$ (isolated nodes) to $\mathcal{O}(N)$: the multiresolution modularity of Reichardt and Bornholdt (RB) \cite{RB}, of Arenas, Fernández and Gómez (AFG) \cite{AFG}, the local fitness method of Lancichinetti, Fortunato and Kertész (LFK) \cite{Lancichinetti}, the Potts model of Ronhovde and Nussinov (RN) \cite{RN}, the Markov autocovariance stability of Delvenne, Yaliraki and Barahona (MAS) \cite{MAS}, the hierarchical likelihood method of Clauset, Moore and Newman (CNM) \cite{newman_nature}, and the Markov Cluster Algorithm of van Dongen (MCL) \cite{MCL}. Naturally, these methods are expected to find the proper community assignments both to Fig. \ref{fig:2cliques} and \ref{fig:sw_2cliques} at some parameter values. However, there is no guarantee that these values are also the proper ones for the rest of the graph. Consequently, it is not clear how a resolution parameter should be set: the natural idea is to find the longest interval of the resolution parameter value in which the community structure does not change, but when the optimal parameter value is different for different regions in the graph, the longest stable interval not necessarily reflects the optimal communities.\\
Furthermore, the fitness values do not help us to tell good clusters from bad ones, like Fig. \ref{fig:2cliques} from Fig. \ref{fig:sw_2cliques}. For most multiresolution methods (RB, AFG, LFK, RN), it is very easy to see that the fitnesses of two clusters are the same given that all nodes has the same in- and outdegrees, independently of the shape of the clusters. Note that it is also true for most single resolution methods. For MAS it is not trivial. Therefore, empirical tests were conducted to check it. According to them, Fig. \ref{fig:2cliques} was found empirically to be at least as good as \ref{fig:sw_2cliques}\footnote{In this case, only 1 link to the rest of the graph was used. Rest of the graph, represented by a single node having self-loops, was assigned 118 edges inside, resulting in a total of $L = 150$ edges. Stability values were calculated from $0.01$ to 100, the step size being $0.01$ below $1.0$ and 1 above.}. Finally, regarding MCL and CNM, they have no fitness function\footnote{CNM does have a fitness function, but it corresponds to a full hierarchical dendrogram, not to any partitions obtained by cutting the dendrogram at some point}, the only accessible quantity about the community structure is the parameter interval in which it is stable.

Finally, there are hierarchical methods, which look for series of smaller and smaller (or larger and larger) clusters hierarchically embedded into the previous ones. Similarly to multiresolution methods, they are expected to contain good clusters in the outputted hierarchy. However, when looking at a graph having a simple one-level community structure, the question how to select the proper levels of the outputted hierarchy arises. The easiest way is to use the lowest level communities. Unfortunately, it is not a reliable procedure, as the lowest-level clusters may be just parts of the communities of the optimal partition or cover (see the Appendix for details). A second idea can be to assign significance scores to the communities on different levels, in the spirit of \cite{stat_signif_Andrea}. Although this approach might reliably qualify the found communities, a new version of statistical significance taking into account the internal cohesion is required. Furthermore, one should be very careful not to impose unnecessary constraints, like prohibiting overlaps, when constructing a hierarchical method.

A further question is whether a method provides information about the shape of the found communities or not. Recent analysis of real-world networks highlights the relevance of this issue \cite{various_comparison}, \cite{mobilephone_comparison}. Several methods are based on simply counting the internal and/or external edges, or degrees at most: LFK, Labelpropagation, Infomap, modularity optimization (and equivalents), Hofman \& Wiggins, Hastings, Ronhovde \& Nussinov, Newman \& Leicht, Wang \& Lai, Bickel \& Chen, Karrer \& Newman, Infomod, Ahn et al., OSLOM. Consequently, they do not see any difference in the distribution of the links, e.g. Fig \ref{fig:2cliques} and \ref{fig:sw_2cliques} get the same fitness values. Only Clique Percolation and Radicchi et al.'s\footnote{If the stopping criterion of their heuristic is considered as part of the definition.} method have some very limited requirement about cohesion built in the definition of communities.

The conclusion is that none of the reviewed methods is able to successfully apply both the separation and the cohesion criterions. They susceptible either to glue together well-separated subgraphs or to overpartition a cohesive subgraph. Future network designs should consider cohesion as well as separation.

\section{Community detection in a two dimensional parameter space}

In this Section, a new method for community detection is introduced. Its main goal is to present a method which takes into account both criterions defined in Sec. \ref{s:def}. First, the LFK method will be reviewed, which will serve as a starting point for the new method. Then, a composite fitness will be constructed which takes into account the separation and cohesion criterions. Finally, a heuristic optimization procedure for the composite fitness will be described, which finds locally dense subgraphs on all scales, and also able to recover hierarchical structures.

\subsection{The LFK method}

The LFK method \cite{Lancichinetti} optimizes the local fitness function
\begin{equation} \label{e:LFK}
 f^C=\frac{K_{in}^C}{(K_{in}^C+K_{out}^C)^{\alpha}}
\end{equation}
where $C$ denotes a subgraph, $K_{in}^C$ and $K_{out}^C$ are the total number of inside and outside degrees in $C$, respectively, and $\alpha$ is a tunable exponent for setting the size scale of the communities to be found. Running the method with large $\alpha$ values result in small clusters, small values in large clusters. The recommended range for $\alpha$ is $0.5$-$2$.

The practical implementation of the optimization works as follows. The communities are found one-by-one, independently of each other. First, a seed node is selected from which the new community will be grown. Then, the node which can best improve the fitness of the cluster is added. This addition is repeated until the fitness reaches a local optimum. After each addition, removal of nodes takes place, if the fitness can be enhanced that way. When the fitness cannot be further increased, the actual subgraph is declared a community. The growth process is repeated for all nodes as seeds, or alternatively, until the found communities cover all nodes in the graph.

Although the resolution parameter $\alpha$ can be tuned continuously, \cite{Lancichinetti} suggested that the relevant community structures should be identified by robustness to changes in $\alpha$, i.e. which have the longest interval for $\alpha$ values without change. Changes in the community structure were detected by monitoring the mean fitness of the communities, evaluated at a reference value $\alpha=1$.

\subsection{Implementing the criterions}

For the separation criterion, the following function will be applied
\begin{equation} \label{e:f_S}
 f_{S}^C = \frac{K_{in}^C}{K_{in}^C+K_{out}^C}
\end{equation}
where $C$ is a subgraph, $K_{in}$ and $K_{out}$ are the sums of in-community and out-community degrees, respectively. This is the fitness of LFK \cite{Lancichinetti}, with the multiresolution parameter being set to one. For detecting hierarchical structures, a different solution will be described. Eq. \ref{e:f_S} clearly focuses on the external separation of the clusters, therefore it is suitable as an implementation of the first criterion of the communities.

For the internal cohesion criterion, a possible solution is to consider the second eigenvalue of the Laplacian matrix of the community. The Laplacian of a graph is the matrix $L=A-D$, where $A$ is the adjacency (or weight) matrix, and $D=\diag(k_i)$ is a diagonal matrix containing the degrees (strengths). Its largest eigenvalue is always 0 (corresponding to the trivial eigenvector $(1,1,\ldots,1)$). The multiplicity of the largest eigenvalue equals to the number of connected components in the graph. This gives the hint that if two distinct graphs are got connected by a single (weak) link, the Laplacian gets only a slight perturbation (compared to the case of two connected components), which splits the double degeneracy of the first eigenvalue, such that a new eigenvalue close to zero appears\footnote{The diffusion matrix was also considered, but it prefers star-like graphs too much.}. In fact, it is known that the second eigenvalue of the Laplacian measures ``how difficult is to split the graph into two large pieces'' \cite{Mohar}.

\begin{figure}[!h]
\begin{center}
\subfloat[]{\label{fig:l2series_1}\includegraphics*[width=2.5cm]{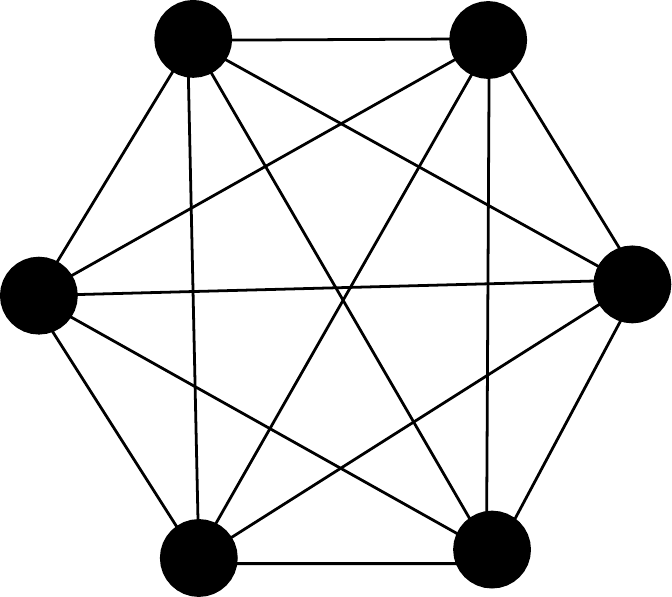}}
\qquad\qquad\qquad
\subfloat[]{\label{fig:l2series_2}\includegraphics*[width=3cm]{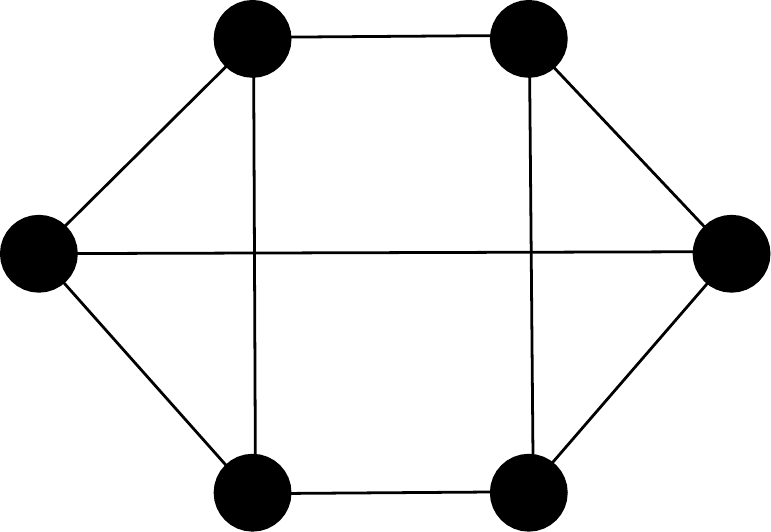}}\\
\subfloat[]{\label{fig:l2series_3}\includegraphics*[width=2.2cm]{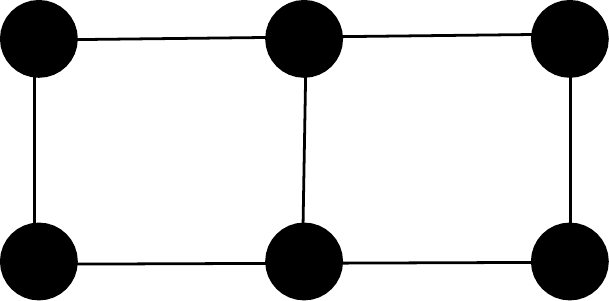}}
\qquad\qquad
\subfloat[]{\label{fig:l2series_4}\includegraphics*[width=5cm]{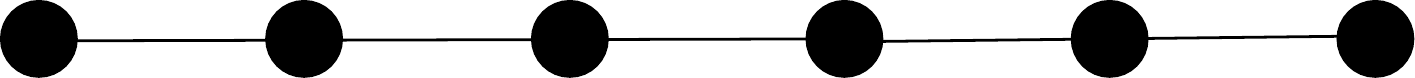}}
\end{center}
\caption{Graphs with different second Laplacian eigenvalues. $\lambda_2^{(\text{a})}=6$, $\lambda_2^{(\text{b})}=2$, $\lambda_2^{(\text{c})}=1$, $\lambda_2^{(\text{d})}=0.268$. The maximal value of $\lambda_2$ is 6 in all cases.}
\label{fig:lambda2_series}
\end{figure}

For some important special cases the second eigenvalue can be calculated:
\begin{itemize}
 \item[-]for full graphs of $n$ nodes (clique), $\lambda_2=-n$
 \item[-]for a star-graph, $\lambda_2=-1$, independently of $n$
 \item[-]for a linear chain, $\lambda_2=-2+2\cos(\pi/n)\to0$ as $n\to\infty$
 \item[-]for two $n$-sized cliques attached by a single link (having weight $\epsilon$) (like on Fig. \ref{fig:2cliques}), $\lambda_2\approx-\frac{2\epsilon}{n+2\epsilon}$, which also goes to 0 as $n\to\infty$.
 \item[-]for a disconnected graph, $\lambda_2 = 0$. This may seem trivial, but most methods give a finite score for disconnected communities; it is not without precedent that such objects can be produced in reality \cite{mobilephone_comparison}. Although this problem can be avoided by a properly designed heuristic of a method, disconnected communities should be punished by definition.
\end{itemize}
Calculation for the two cliques is in the Appendix, other results can be found in \cite{Fiedler}. These cases confirm that the second eigenvalue is useful for quantifying the cohesion criterion of the definition of communities. For an illustration, on Fig. \ref{fig:lambda2_series} a few example graphs with their second Laplacian eigenvalues are shown.

The separation fitness term $f_S$ ranges from zero to one. In order to compose it together with the cohesion fitness, the latter should also be in the interval $[0,1]$. Therefore, $\lambda_2$ needs some transformations before application as fitness.\\
As can be seen from the above examples, for the worst cases $|\lambda_2|$ is of the order of $1/n$, therefore the lowest point of the $|\lambda_2|$-scale will be set to $1/n$. The highest point is trivially given by $n$. It is reasonable to assume that most subgraphs have $|\lambda_2|=\textrm{o}(|C|)$. Furthermore, several subgraphs can have worse internal cohesion than the star graph, thus having $|\lambda_2| \in [0,1]$. To take into account these effect, $\log |\lambda_2|$ will be more useful than $\lambda_2$. So, in order to obtain a quantity between 0 and 1, the minimum will be subtracted and divided by the maximum,
\begin{equation} \label{eq:lambda2}
\begin{split}
f_{C}^C=&\frac{\log |\lambda_2| - \log 1/|C|}{\log |C| - \log 1/|C|} =      \frac{1}{2}+\frac{1}{2}\frac{\log|\lambda_2|}{\log|C|}, \quad \text{if } |C|>1\\
		 =&\,0\quad\qquad\qquad\qquad\qquad\qquad\qquad\quad\quad\;\;\,\;\;\: \text{if }|C|=1
\end{split}
\end{equation}
where $|C|$ is the number of nodes in the community. The above measure happens to be $0.5$ if $\lambda_2$ is 1, e.g. for the star-graph. I wish to emphasize that eq. \ref{eq:lambda2} is only one possible proposition for taking into account the internal cohesion, although a promising one -- better measures may exist. The same is true for the choice of $f_{S}$.

The cohesion fitness $f_C^C$ opens the way for constructing tests assessing the performance of community detection methods regarding the cohesion of the found communities. One may generate a graph with built-in communities which separation is controlled, like in the LFR benchmark \cite{SFbenchmark1}, then randomly select pairs of clusters and increase the interconnection between the two members of each pairs to some predefined value, finally calculating $f_C^C$ of the pairs. Running the detection method and measuring the ratio of pairs not split as a function of $f_C^C$ may indicate how strongly focuses the method on cohesion.

The next question is how to combine $f_{separation}^C$ and $f_{cohesion}^C$. Thinking in a two dimensional space of $f_{S}^C$ and $f_{C}^C$, a natural approach is to get as far from the point $(0,0)$ as possible. This implies
\begin{equation} \label{e:fullfitness}
f^C=\sqrt{(f_{S}^C)^2+(f_{C}^C)^2}
\end{equation}
so the fitness is the euclidean distance from $(0,0)$. Again, this is just one possibility, better combinations may exist. E.g. the relative weight of $f_{S}$ and $f_{C}$ may be adjusted in a more well-grounded way. However, eq. \ref{e:fullfitness} is able to pass the test raised by Fig. \ref{fig:counterexample}: for Fig. \ref{fig:2cliques}, $\lambda_2^{2\text{cliques}}=0.258$, $f_C^{2\text{cliques}}=0.228$, $f^{2\text{cliques}}=0.995$ while for a single clique $\lambda_2^{1\text{clique}}=6$, $f_C^{1\text{clique}}=1$, $f^{1\text{clique}}=1.371$. For Fig. \ref{fig:sw_2cliques}, $\lambda_2^{12\text{nodes}}=3.268$, $f_C^{12\text{nodes}}=0.738$, $f^{12\text{nodes}}=1.218$, and for the best subgraph, a triangle, $\lambda_2^{\text{triangle}}=3$, $f_C^{\text{triangle}}=1$, $f^{\text{triangle}}=1.077$.

Beyond enabling one to decide whether a given subgraph is a community or not (by requiring local optimality), the above definition makes it possible to assess how good community it is. This is also possible with another definitions, e.g. by using the modularity function, but here, communities are placed on a 2-dimensional space instead of 1 dimension. This gives rise to an interesting possibility for characterizing the communities, like ``very cohesive but densely connected outwards'' or ``well-separated but poorly interconnected''. Considering Fig.\ \ref{fig:2cliques}, one may think that the latter is not really a community. But for large subgraphs, it may make sense to consider a well-separated subgraph as a community, as common sense says that large communities should be looser than small ones.

\subsection{Community detection in reality}

In this section, the details of practical implementation of the new method are discussed. Most importantly, in order to actually find the communities, a heuristic carrying out the optimization of eq. \ref{e:fullfitness} is needed. Furthermore, there is a second problem of detecting communities hierarchically embedded into each other. These two questions will be answered by a common solution.

The heuristic is based on the one of the LFK method  \cite{Lancichinetti}. Among its details, the LFK heuristic contains a tunable parameter (denoted as $\alpha$), which is claimed to be able to recover communities at different hierarchical levels. Lowering this parameter $\alpha$ results in increased community sizes. Hierarchical levels are supposed to be stable against the variation of $\alpha$, so there should be long intervals for $\alpha$ for which the communities do not change. However, large graphs may lack long stable intervals, as some changes occur around any parameter value (data not shown). Therefore, a new method for investigating hierarchical structure is needed. I dropped the idea of using threshold values of $\alpha$, corresponding to community structures at different scales, which should be simultaneously valid for all communities, and I will treat each community separately.

Similarly to \cite{Lancichinetti}, each community is grown from a seed node. It is important to note that each seed node can result in a series of (successively larger) communities. Growth consists of successively including the neighboring node which increases most the fitness defined by eq. \ref{e:fullfitness}. When there is no neighboring node which inclusion can improve the fitness, the stage of node removal begins. Here, the fitness of the cluster is tried to be improved by excluding nodes from it (with the exception of the seed node, which is not permitted to be excluded). It finishes when no further removal can improve the fitness. Then, growth begins again, if possible. The grow-shrink cycle is iterated, as long as the fitness can be improved. When no improvement is possible (there is a local optimum of the fitness), the actual list of nodes is registered as a valid community. After that, the algorithm tries to find a larger community, which contains the current one. This way, hierarchical structures can be revealed. In order to do it, first the growing cluster should escape from the basin of attraction of the current local optimum. Therefore, the cluster is forced to grow, by successively including the neighboring nodes which decrease the fitness the least. After some steps of forced growth, when increasing the fitness becomes again possible, the algorithm turns back to the normal grow-shrink procedure, until a new local optimum is found, signing a new community. The cluster keeps hopping from local optimum to another local optimum until it grows so large that it contains the whole graph. Then a new growth process starts from a new seed node. At the end of its growth process, it includes the whole graph again, unless it encounters a local optimum which has been already found, i.e. the corresponding community has already been registered. In this case, the growth process is stopped. Then, another growth process starts from a not-yet-used seed node. In contrast to \cite{Lancichinetti}, all nodes in the graph are used as seed nodes, in order not to miss good communities. When the growth process beginning from the last seed node finishes, the algorithm ends, and the registered communities are written to the output.\\
There are a few additional tricks. First, if escaping from a local optimum seems to be hard, i.e. after changing from forced growth to the normal grow-shrink stage we still end up in the previous local optimum, the cluster is restored to the state where it had its maximal size (the beginning of one of the removal sessions), then 2 steps of forced growth is applied before the normal grow-shrink cycle begins. A second trick is that when judging the identity of two communities, they are considered identical if at least $80\%$ of the larger community is a subset of the smaller one\footnote{If the criterion were based on some percent of the smaller group, subset-superset pairs would be considered identical.}. In case of identity, the community which has the higher fitness is kept in the registry.\\
The algorithm, although based on the one of \cite{Lancichinetti}, differs in several points: from one seed, several communities can be reached instead of only the smallest one; node removal occurs when node addition is not possible instead of after each addition (this trick also speeds up the algorithm); seed node is not permitted to be removed; all nodes are used as seeds instead of the not-yet-covered nodes. An algorithm similar in spirit was described in \cite{clauset}.
The results in the next section are obtained using this method, unless stated otherwise explicitly. The software realizing the algorithm is available at\\ \verb+http://www.phy.bme.hu/~tibelyg/+.

\section{Test results}

\begin{figure}[!h]
\begin{center}
\includegraphics*[height=9cm]{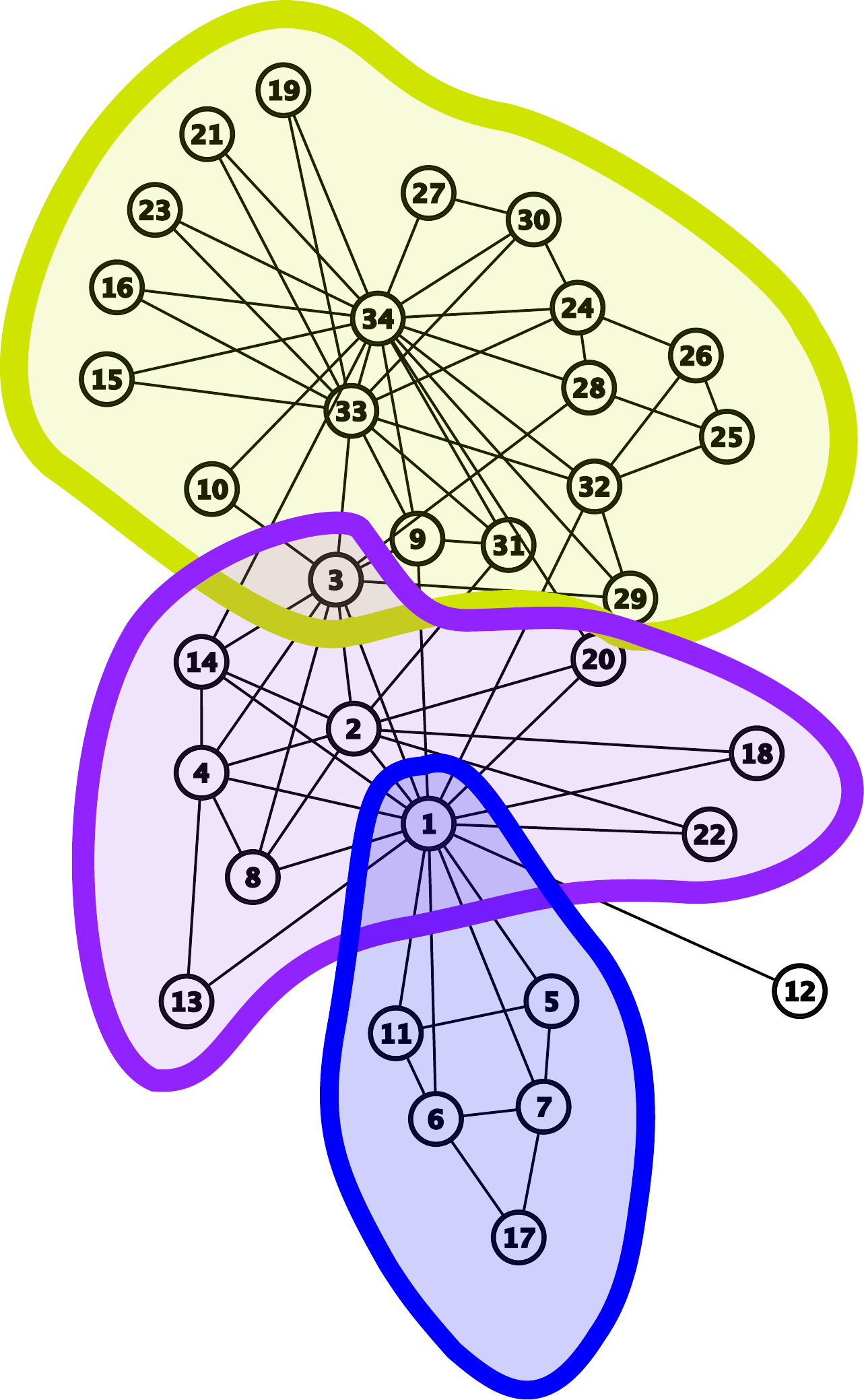}
\phantom{a}
\includegraphics*[height=9cm]{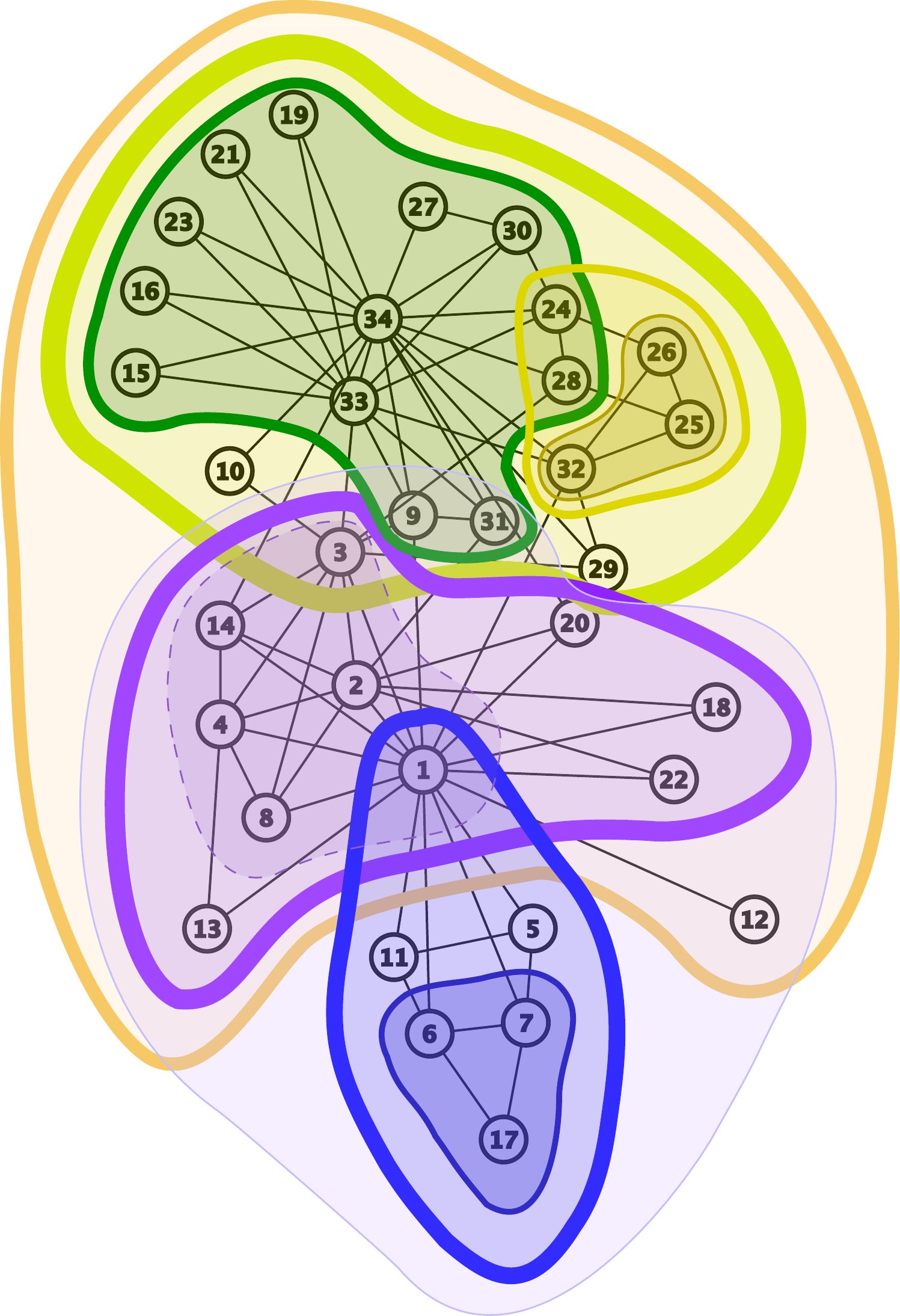}
\end{center}
\caption{(Color online) The 3 (left) and 10 (right) best found communities of the Zachary karate club. On the right, thicknesses of lines indicate the ordering of the statistical significance values (running from $0.002$ to $0.42$, plus $0.81$ for the dashed line-bordered community). Note that node 12 is contained only by large communities.}
\label{fig:zach}
\end{figure}

Probably the most frequently used test is Zachary's karate club friendship network\cite{Zachary source}. Due to a dispute between two prominent persons (node 1 and 34), the club split into two during sociological observation, and the memberships in the new clubs are known. As the split occurred more or less along a border of two visible communities, new community detection algorithms are usually claimed to pass the test if they reproduce the split. However, the aim is the detection of \emph{topological} modules, not functional ones, so the result of the sociological study is not a strict criterion for judging the output of any community detection method. E.g., node 10 has $1-1$ links to each of the new clubs, so ``misplacing'' it (compared to the split) may not be considered as a fault. Or node 12, which attaches only to node 1, is hard to be considered as part of a ``densely interconnected'' cluster.\\
The algorithm finds $33$ groups, containing several non-relevant ones, like pairs of nodes. Therefore, a filtering procedure is required. The statistical significance of the resulting communities \cite{stat_signif_Andrea, OSLOM} is utilized for this purpose. The statistical significance can be sensitive for missing nodes \cite{stat_signif_Andrea}, therefore each cluster is allowed to be completed with the neighboring node which optimizes the statistical significance. Then the clusters are ordered according to their statistical significance. The first 3 clusters provide a single-level community structure, corresponding to 3 known communities, with 2 overlapping and 1 homeless nodes (Fig. \ref{fig:zach}, left panel). Taking a look at the subsequent clusters provides information about the multi-scale structures in the graph. The next few clusters reveal cluster cores and hierarchical decomposition of the network (Fig. \ref{fig:zach}, right panel). The statistical significance score is quite capable of distinguishing meaningful structures; there is a gap between $0.42$ and $0.81$, so setting a threshold to $0.5$ selects the multi-scale clusters which would be approved by a human investigator. There is only one exception, the almost-full-clique of nodes \{1, 2, 3, 4, 8, 14\} has significance $0.81$, which is probably the consequence of neglecting the internal cohesion by the current form of statistical significance.

\begin{figure}[!h]
\begin{center}
\includegraphics*[height=6cm]{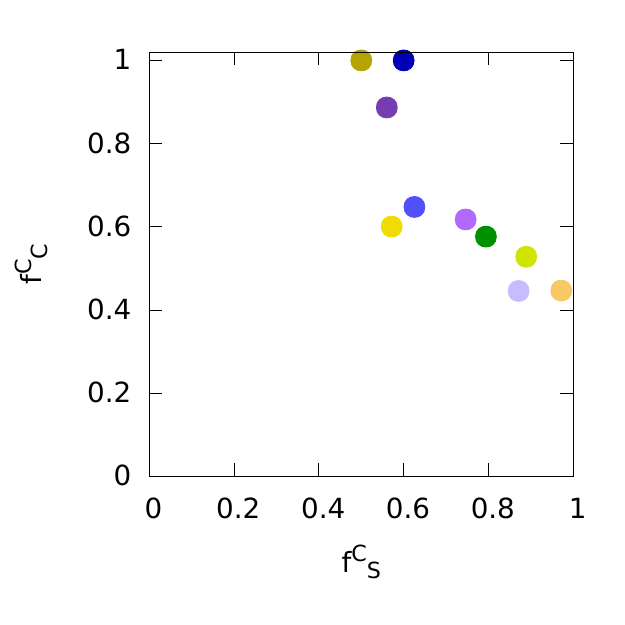}
\end{center}
\caption{(Color online) Positions of the Zachary communities on the $f_{S}$-$f_{C}$ plane. Small groups tend to cluster at North, and large groups at East.}
\label{fig:zach_f1-f2}
\end{figure}

\begin{figure}[!h]
\begin{center}
\subfloat[]{\label{fig:SFtest-a}\includegraphics*[width=0.47\textwidth, trim=0.4cm 0cm 0.7cm 0cm, clip]{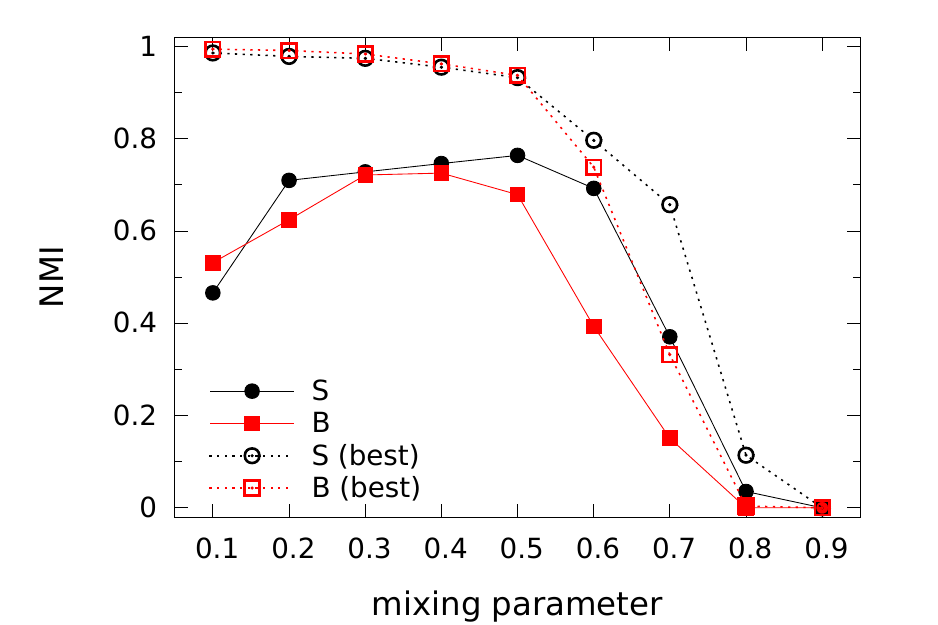}}
\hspace{0.5cm}
\subfloat[]{\label{fig:SFtest-b}\includegraphics*[width=0.47\textwidth, trim=0.4cm 0cm 0.7cm 0cm, clip]{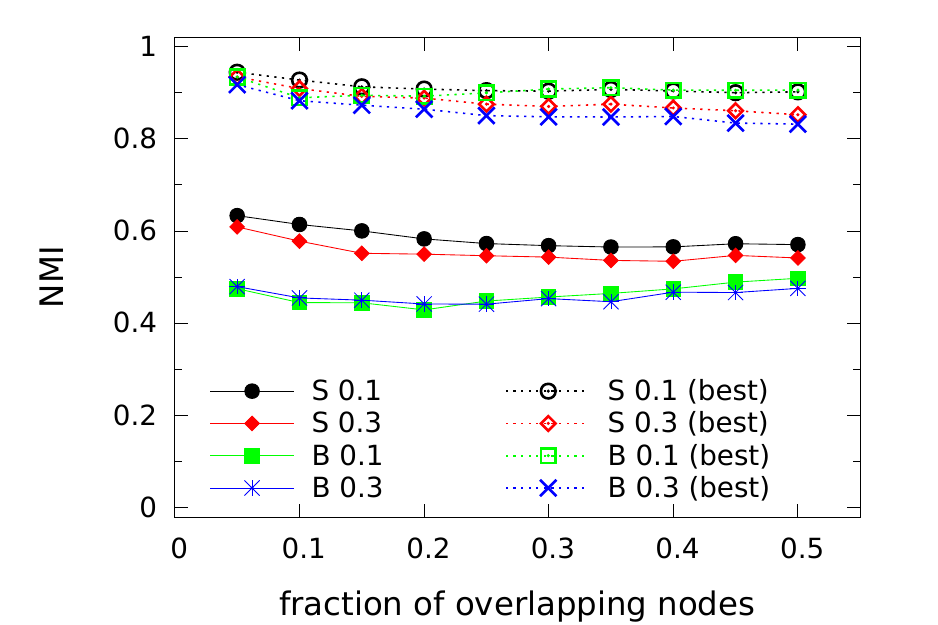}}
\end{center}
\caption{(Color online) Results on the LFR benchmark. Panel (a) corresponds to unweighted, undirected and non-overlapping tests, while panel (b) corresponds to overlapping tests. Overlapping tests were done at two different values of the mixing parameter, at $\mu = 0.1$, $0.3$. For both panels: full symbols and lines correspond to the applied filtering, empty symbols with dotted lines correspond to perfect output filtering.}
\label{fig:SFtest}
\end{figure}

The currently most advanced class of benchmarks was introduced by \cite{SFbenchmark1}. In these so-called LFR benchmarks, the network size and edge density are freely adjustable, and more importantly, the node degrees and the community sizes are distributed according to power-law distributions, with tunable exponents. Communities are defined through a prescribed ratio of inter-community links for each node (mixing ratio, $\mu$), similarly to the preceding GN benchmark class \cite{Q}. Generalizations for weighted and directed networks, and for overlapping communities also exist \cite{SFbenchmark2}.\\ A wide-scale comparison of different community detection methods using the LFR benchmark was done by \cite{comparison}. For the ease of comparison, the parameter values of \cite{comparison} are applied here: the networks consist of 1000 nodes, the average degree is 20, the maximal degree is 50, the exponent of the degree distribution is -2 and the exponent of the community size distribution is -1. There are two types of networks, for the S type the community sizes are between 10 and 50 (``small'') and for the B type they are between 20 and 100 (``big''). In \cite{comparison}, networks of 5000 nodes were also investigated. Due to the large computational time, they are omitted here\footnote{it does not mean that a single 5000-sized graph is too large, however, a few hundred of them are}. Also for computational time considerations, the detecting algorithm stopped growing the communities over a predefined size, 120 for the S case and 220 for the B case. All measurement values are obtained from runs on 10 different networks.\\
Similarity of the built-in and the obtained community structures are quantified by a variant of the normalized mutual information (NMI), which is able to handle overlapping communities \cite{Lancichinetti}. This is the similarity measure applied by \cite{comparison}\footnote{Both the LFR benchmark and the generalized normalized mutual information are freely available from the authors' websites, \texttt{http://sites.google.com/site/santofortunato/inthepress2} and \texttt{http://sites.google.com/site/andrealancichinetti/software}}.\\
Selecting the most relevant communities from the abundant output was done similarly to the previous case. The clusters were completed by 1 neighboring node, if that improved the statistical significance, and sorted with respect to the statistical significance scores. The clusters containing at least 1 uncovered node were accepted one by one until all nodes were covered.\\
To see the potential of the new method, and check the effect of the output-filtering, the communities corresponding best to the built-in original ones were also selected from the algorithm's output. The results are plotted on Fig. \ref{fig:SFtest} (a). The filtered results are similar to the ones of the lower performing algorithms in \cite{comparison}, while optimal selection provides much better scores, although still not as good as the best methods. 
The large difference between the optimal and the statistical significance-based results is quite surprising, especially in the light of the fact that statistical significance in itself is able to provide excellent results on the LFR benchmark \cite{OSLOM}.\\
The algorithm was also tested on networks with overlapping communities. In this case, clusters having significance score below $0.1$ were accepted, similarly to \cite{OSLOM}. Fig. \ref{fig:SFtest} (b) shows that the effect of the imperfect output-filtering is again very large, an ideal selection scheme would allow very good results. This is not surprising, as other algorithms based on the one of \cite{Lancichinetti} also give excellent results on overlapping communities \cite{GCE}.\\

\begin{figure}[!h]
\begin{center}
\includegraphics*[height=10cm]{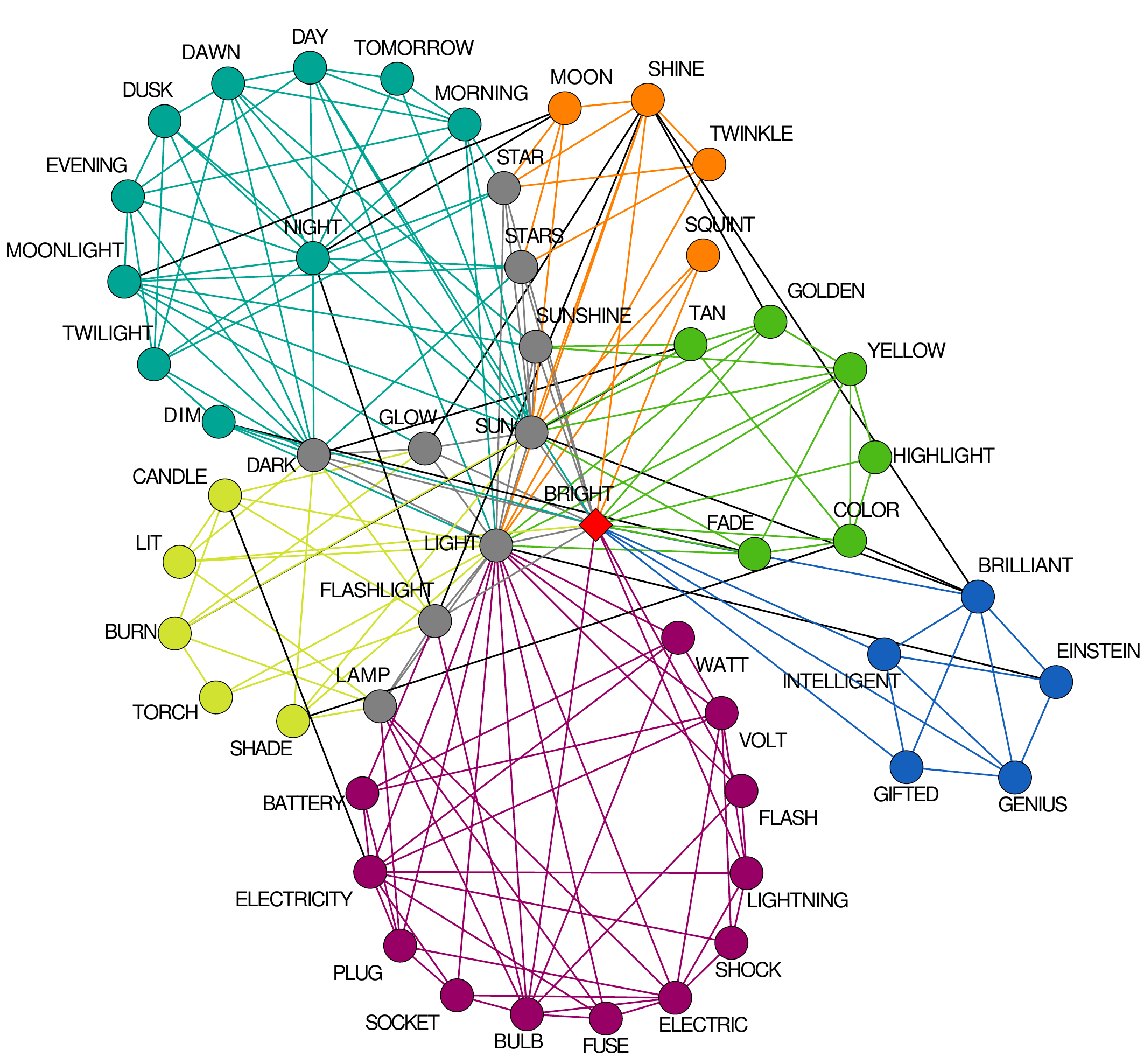}
\end{center}
\caption{(Color online) Communities around \emph{bright}, on the first hierarchical level. Color denotes communities. Gray shows overlapping nodes and edges. Black edges are between different communities.}
\label{fig:wa}
\end{figure}

Finally, the new method was applied to a word association graph built from the University of South Florida Free Association Norms \cite{wa}. Here, nodes are words and edges show that some people associated the corresponding two words. The network has 5018 nodes with mean degree $\langle k \rangle = 22.0$. It is a frequently used example of overlapping community structure \cite{OSLOM}, \cite{CPM}. Although edge weights are accessible, the algorithm was applied to the unweighted version of the network. As an illustration, low-level communities around the word \emph{bright} are plotted on Fig. \ref{fig:wa}. An interesting effect is the appearance of \emph{overlapping edges}, due to the heavy overlap in the network.

In conclusion, although selecting the relevant communities from the output is not an already solved task, the algorithm gives good results on the Zachary karate club, and performs reasonably on the LFR benchmarks. It should be noted however, that due to the internal cohesion criterion, this algorithm's output is not intended to perfectly match benchmarks like GN and LFR, which define communities solely on the basis of external separation. An additional observation is reported here: on GN benchmark graphs\footnote{results are omitted, as the presented LFR benchmark is a generalization of the GN.} with nodes having exactly the prescribed in- and out-degrees, at large mixing ratios communities deviating from the built-in ones but having better-than-designed mixing ratios were found. Note that the new method does not optimize just for external separation, so even better ``spontaneous'' communities may exist. This phenomenon, although not being a huge surprise, raises the question how to judge precisely a community detection method's output at large mixing ratios, as the known community structure may not be trusted to 100\%.

\section{Discussion \& Conclusions}

An important aspect of all community detection methods is the \emph{running time}. In the case of the new method described above, the time requirement is as follows. Starting a new community from each node contributes a factor of $N$ to the CPU time. Evaluating the eigenvalues of a community $C$ plus one extra node takes $2/3\:(|C|+1)^3$. Assuming that $C$ has $\text{const}\cdot\langle k\rangle\cdot|C|$ neighboring nodes (i.e., on average, each node has a constant fraction of its neighbors outside $C$), running time can be estimated as 
\begin{equation}
T \approx N\cdot\sum_{|C|=1}^{|C|_{\text{max}}} \text{const}\cdot\langle k\rangle\cdot|C|\cdot\;(|C|+1)^3 \approx N\cdot\text{const'}\cdot\langle k\rangle\cdot|C|_{\text{max}}^5
\end{equation}
A naive estimate for $|C|_{\text{max}}$ would be $N$. However, as more and more community growing processes finish, the newly started communities are expected to terminate in a previously discovered community earlier and earlier, on average. Of course, some communities will reach $|C|=N$. Therefore,
\begin{equation}
 T \propto N^{5+\delta},\qquad \delta \in[0,1]
\end{equation}
which is huge and clearly denies the analysis of even medium-sized graphs ($\mathcal{O}(10^4)$ nodes) without further improvements. Note that graphs of thousands of nodes may be manageable, like the word association graph shown above, which took 56 hours on a single CPU. One possibility is to choose the initial seed more intelligently, starting communities from promising seeds. \cite{GCE} achieved good results in this aspect. An intelligent seed selection is also important if the number of communities in a cover is larger than $N$, or if some communities have only overlapping nodes -- in this case, it may happen that all growth processes miss a certain community.\\
Other important question is the applicability of an advanced eigenvalue solver. Arpack++ \cite{arpack++} and SLEPc \cite{slepc} were tried. The experience was that -- despite their good asymptotic performance in the large matrix limit -- for the occurring several small subgraphs the overhead of these complicated machineries was so large that made the final running time much higher than those obtained with the QR-decomposition algorithm.
\\
Doing optimization in a multi-parameter space is a nontrivial task, because different parameters can lie in different ranges. Therefore, an important direction for future research is to investigate the best combination of the parameters in the fitness function, based on the evaluation of empirical data.
\\
Finally, filtering the relevant communities from the found ones is also a challenging task. The natural approach is to apply statistical significance, which should be applied even if filtering was not needed. However, deciding the threshold significance value is not necessarily trivial in all cases. Furthermore, the current form of statistical significance accounts only for the separation of the community, not for its internal cohesion. This manifests itself e.g. in the low score of the almost-full-clique subgraph in the Zachary karate club (the dark purple group on Fig. \ref{fig:zach}). As the main advantage of the fitness function of eq. \ref{e:fullfitness} is the inclusion of cohesion, it would be important to develop a statistical significance taking it into account. 

\emph{Conclusions}. The community detection problem currently suffers from two fundamental deficiencies. First, there is no definition of community which is precise enough to allow constructing community finding methods. Second, thorough testing a proposed algorithm is problematic, not independently from the previous difficulty. I attempted to improve both issues.

In this paper, I proposed a formal list of required properties for locally dense subgraphs, taking a step towards an applicable definition of the term ``community''. Two properties, external separation and internal cohesion (``shape'') were named. External separation has already been applied by some of the community detection methods, and also by benchmarks. Internal cohesion was not considered explicitly earlier. No current method was found which satisfactorily applies both criterions. I demonstrated on simple examples that both properties are necessary; discarding either of them leads to counterintuitive results. Beyond allowing to construct new methods, these two criterions can also be used as a basis for testing existing ones. They also allow the characterization of a community by two independent quantities, instead of a single scalar.

I proposed a new composite fitness function which takes the two criterions into account. For the quantification of the internal cohesion, the second eigenvalue of the Laplacian matrix is applied, which provides appropriate results on characteristic graphs like cliques or chains. I also proposed a heuristic, by redesigning the LFK heuristic \cite{Lancichinetti}, which can find overlapping locally dense subgraphs of all scales, producing much less output than multiresolution methods but with less restrictions than imposed by assuming a hierarchical structure. Runs on the Zachary network and LFR benchmarks showed that the method is able to provide the expected results. Overlapping communities can be detected especially efficiently, similarly to other LFK-based heuristics \cite{GCE}. However, significant improvements are yet to be implemented; e.g. reducing the running time, finding a more effective filtering procedure for the output, or fine-tuning the relative weight of the separation and the cohesion terms in the fitness function.

\section{Acknowledgments}

I wish to thank János Kertész for several useful suggestions. I also thank the Eötvös University for the access to its HPC cluster, and Andrea Lancichinetti, who proposed the Arpack++ package and was extremely helpful about his software. Thanks are due to the authors of the OSLOM \cite{OSLOM}, LFR benchmark \cite{SFbenchmark1}, overlapping mutual information \cite{Lancichinetti}, Radatools \cite{radatools} and linegraph-creator \cite{EvansLambiotte} softwares for making their code publicly available. I am grateful for the referees for several comments which significantly improved the manuscript. Financial support from EU's 7th Framework Program's FET-Open to ICTeCollective project no. 238597 is acknowledged.

\newpage

\section*{Appendix}

\appendix

\section{Second eigenvalue of two weakly connected cliques}

Assume two cliques of $n$ nodes, edge weights are 1. The two cliques are attached by a single edge having weight $\epsilon$. Then the eigenvalue equations for the Laplacian matrix are
\begin{align}
\label{e:1}  &\sum_{\substack{j<n\\ j \neq i}} x_j + x_n - (n-1+\lambda)x_i=0 \qquad \forall i<n\\
\label{e:2}  &\sum_{\substack{k>n\\ k \neq i}} x_k + x_{n+1} - (n-1+\lambda)x_i=0 \qquad \forall i>n+1\\
             &\sum_{j<n} x_j + \epsilon \cdot x_{n+1} - (n-1+\epsilon+\lambda)x_i=0 \qquad i=n\\
             &\sum_{k>n+1} x_k + \epsilon \cdot x_n - (n-1+\epsilon+\lambda)x_i=0 \qquad i=n+1
\end{align}
Adding the last two equations gives
\begin{equation}
 \sum_j x_j - x_n - x_{n+1} + \epsilon x_n + \epsilon x_{n+1} - (n-1+\epsilon+\lambda)x_n - (n-1+\epsilon+\lambda)x_{n+1} = 0
\end{equation}
The eigenvector corresponding to the first eigenvalue (which is zero) is the constant vector, therefore for all other eigenvectors the sum of components should be zero in order to be orthogonal to the first one. Consequently $\sum_j x_j=0$. Applying this and a minimal algebra results
\begin{align}
 x_n(\lambda+n)+x_{n+1}(\lambda+n)&=0\\
 x_n&=-x_{n-1} \qquad \text{if } \lambda \neq -n
\end{align}
If $\lambda=-n$ then eqs. \ref{e:1}-\ref{e:2} reduce to $\sum_{j\leq n}x_j=0$ and $\sum_{k \geq n+1}x_k=0$. Now consider the eigenspace corresponding to $\lambda=-n$, and look for eigenvectors such that $x_n=x_{n+1}=c$, $\sum_{j<n}x_j=-c$, $\sum_{j>n+1}x_j=-c$. In this eigenspace the number of free parameters are $1+2\cdot(n-2)$, corresponding to $c$ and $x_1 \ldots x_{n-1}, x_{n+2} \ldots x_{2n}$ with two constraints. Altogether the dimension of the eigenspace (the multiplicity of $\lambda=-n$) is $2n-3$. Adding the $\lambda=0$ case, we are left with at most two unknown eigenvalues.

For $\lambda \neq -n$, we look for the solutions in the form $(a,\ldots,a,b,-b,-a, $N$\ldots,-a)^T$. Then the eigenvalue equations are
\begin{align}
 (n - 2) a + b - (n - 1 + \lambda) a &= 0\\
 (n - 1) a - \epsilon \cdot b - (n - 1 + \epsilon + \lambda) b &= 0
\end{align}
After simplifications,
\begin{align}
 - (1 + \lambda) a + b &= 0\\
 (n - 1) a - (n - 1 + 2\epsilon + \lambda) b &= 0
\end{align}
Expressing $\lambda$ from these equations reads
\begin{align}
 & \lambda = \frac{b}{a}-1  \label{e:3}\\
 & \lambda = - (n - 1 + 2\epsilon) + (n-1)\frac{a}{b}
\end{align}
Writing $\lambda=\lambda$ results
\begin{align}
 & -n+1-2\epsilon+(n-1)\frac{a}{b}=\frac{b}{a}-1\\
 & (-n+1-2\epsilon)\frac{b}{a}+(n-1)=\left(\frac{b}{a}\right)^2-\frac{b}{a}
\end{align}
Introducing $x=b/a$ gives
\begin{align}
 & -x^2 +(-n+2-2\epsilon)x+(n-1)=0\\
 & x_{1,2}=\frac{n-2+2\epsilon\pm\sqrt{(n-2+2\epsilon)^2+4(n-1)}}{-2}
\end{align}
The term under the radical symbol can be approximated using the first two terms of the Taylor series $\sqrt{1-x}\approx1-x/2$
\begin{align}
 \sqrt{\ldots}&=\sqrt{(n+2\epsilon)^2\left(1-\frac{8\epsilon}{(n+2\epsilon)^2}\right)}\approx\\
 &\approx(n+2\epsilon)\left(1-\frac{4\epsilon}{(n+2\epsilon)^2}\right)
\end{align}
which gives
\begin{align}
 x_1&\approx1-\frac{2\epsilon}{n+2\epsilon}\\
 x_2&\approx1-(n+2\epsilon)+\frac{2\epsilon}{n+2\epsilon}
\end{align}
which, using eq. \ref{e:3}, leads to
\begin{align}
 \lambda_1&\approx-\frac{2\epsilon}{n+2\epsilon}\\
 \lambda_2&\approx-(n+2\epsilon)+\frac{2\epsilon}{n+2\epsilon}
\end{align}
meaning that the last two eigenvalues of the Laplacian are found.

\newpage

\section{Review of current methods}

Here, a one-by-one review of methods follows, from the point of view of the separation \& cohesion criterions.

\subsection*{Separation-targeted methods}

\noindent\textbf{Method of Lancichinetti et al. (LFK) \cite{Lancichinetti}} -- although being a multiresolution method, it is informative to take a look at it with the resolution parameter (see eq. \ref{e:LFK}) fixed at $\alpha=1$. Then the fitness function of a community, which is to be optimized, is simply the sum of in-degrees divided by the sum of degrees of the community members. Thus, this method is a clear implementation of the separation criterion. Consequently, it is not sensitive to the internal distribution of edges (Fig. \ref{fig:2cliques} and \ref{fig:sw_2cliques} get the same fitness). The cohesion criterion is absent, so one clique on Fig. \ref{fig:2cliques} has lower fitness than the union of the two cliques. 
\\
\textbf{Labelpropagation \cite{labelprop}} -- the communities are defined as sets of nodes such that every node should belong to the community to which the majority of their neighbors do. Labelpropagation does not qualify the communities, just finds partitions obeying the majority rule. Consequently Fig. \ref{fig:2cliques} can be judged as a proper single community, and Fig. \ref{fig:sw_2cliques} can be split by collecting each second node to the same cluster. 
\\
\textbf{Infomap \cite{Infomap}} -- Infomap aims to minimize the length of the description of a random walk, using clusters. The best description length corresponds to the best trade-off between small cluster sizes (understood in in-degrees) and few links between clusters. It is straightforward to calculate that for the configuration on Fig. \ref{fig:2cliques}, Infomap will properly separate the two cliques unless the number of inter-community links is larger than $6.9\cdot10^7$. Although this resolution limit looks practically unimportant, shows that Infomap has some conceptual problems. If 3 edges are placed instead of 1 between the 2 cliques on Fig. \ref{fig:2cliques}, Infomap will merge the two cliques if the number of inter-community edges in the rest of the network is larger than $149$, which is more than 5 orders of magnitude smaller than the previous threshold. Two consequences should be drawn: Infomap is quite sensitive to the number of inter-community edges, and, as a consequence, it can produce counterintuitive communities in realistic graphs.
\\
\textbf{Clique Percolation Method (CPM) \cite{CPM}} -- communities are defined as maximal sets of adjacent $k$-cliques, $k$ being a parameter. Adjacency holds if $k-1$ nodes are shared by two cliques. Although CPM enforces a very strong cohesion locally, it applies only to $\mathcal{O}(1)$-sized subgraphs of communities. Consequently, there are no cohesion requirements on the scale of the whole community. E.g., the cliques of a cluster might form a chain and the method gives no information about the shape of the cluster. Considering Fig.\ \ref{fig:2cliques}, it is trivial to modify it such that CPM merges the two large cliques into a single cluster, e.g. using 3-cliques. Furthermore, the absence of a single percolating series of neighboring cliques means that a subgraph will not appear as a single community, regardless of its other parameters (see e.g. Fig. \ref{fig:sw_2cliques} applying 4-cliques). Finally, CPM uses the same clique size for the whole network, regardless of local variations in edge density. 
\\
\textbf{Method of Radicchi et al. \cite{Radicchi}} -- there are two possible criterions for communities to choose from: either all community members or only the whole community should have more links inside than outside. Proper communities are found by iteratively bisecting the network, until no bisection can be carried out without violating the criterion used. So, the effective definition is that a community is a subgraph obeying one of the criterions mentioned above \emph{such that no bisection of it can result proper communities}. Fig. \ref{fig:sw_2cliques} with a minor tweak would be split even using the strong definition, assigning every second node to the same community. The tweak is to place the 2 outside links on the $k_{in}=6$ nodes.
\\
\textbf{Method of Estrada and Hatano \cite{Estrada}} -- as it relies on the eigenvalues and eigenvectors of the whole graph, it is a global method. Therefore whether a set of nodes is judged to be a cluster or not depends also on the rest of the graph. Unfortunately, the behavior of the eigenvalues and eigenvectors of the adjacency matrix of a graph are not well understood. Consequently, empirical tests were conducted. If the method is run on only the 12 nodes of Fig.\ \ref{fig:counterexample}, configuration a) is cut into the two proper sets, but configuration b) is cut into several small (overlapping) clusters, such that all triangles form one. When the 12 nodes are attached to a 100-node ring, in which first and second neighbors on both sides of a node are attached to the node (degrees are 4), then for configuration a) the two clusters expand to the first neighboring nodes in the ring, and configuration b) has the same clusters as in the fully separated case. So, if the rest of the graph is not denser than the set of nodes under investigation, it seems that internal cohesion does matter, however external separation not. If the 100-node ring is two degrees denser (first 3 neighbors are attached, degrees are 6), the 12 nodes coalesce into 1 cluster both for configurations a) and b), incorporating a few nearby nodes from the large ring (8 for a) and 6 for b)). For even denser 100-node rings, the 12 nodes become part of a large cluster containing many nodes from the large ring. So, in conclusion, the global character of the method makes it indefinite concerning its behavior to the configurations on Fig. \ref{fig:2cliques} and \ref{fig:sw_2cliques}.

\subsection*{Stochastic blockmodels and spin-based methods}

\noindent\textbf{Modularity optimization \cite{Q}} -- for each community, modularity counts the inside links and their expected values, based on the degrees of the nodes. Due to the well-known resolution limit problem \cite{reslim, reslim2}, the optimal modularity merge the two cliques on Fig. \ref{fig:2cliques} for sufficiently large graphs.
\\
\textbf{Laplacian spectral algorithm by Donetti and Mu\~noz \cite{DM}} -- although the method produces candidate partitions using the spectrum of the Laplacian matrix, the partitions are evaluated using modularity. Consequently, it is equivalent to modularity optimization using a special heuristic, implying all the drawbacks of modularity.
\\
\textbf{Link partitioning method of Evans and Lambiotte \cite{EvansLambiotte}} -- partitioning is done on the so-called line graph, which nodes correspond to the edges of the original graph, and links are drawn between edges sharing a node in the original graph. Variants of the modularity function are proposed as goal function for the partition. Different variants use different weighting schemes of the edges including the addition of self-loops. As these goal functions are still based on counting intra-community edges and subtracting some expected value, the resolution limit problem should appear for large enough graphs.
\\
\textbf{Method of Ronhovde and Nussinov (RN) \cite{RN}} -- it proposes a Hamiltonian $\mathcal{H(\{\sigma\})}=-1/2\sum_{i \neq j} (a_{ij}A_{ij}-\gamma b_{ij} (1-A_{ij}))\delta(\sigma_i, \sigma_j)$, $\mathbf{A}$ being the adjacency matrix, $a_{ij}$ and $b_{ij}$ being edge weights. The configuration corresponding to the minimal Hamiltonian is used as the solution.\\
The Hamiltonian optimizes simply for the edge densities inside clusters (distorted by the $\gamma$ resolution parameter), which tends to be the largest for cliques. Consequently Fig. \ref{fig:sw_2cliques} worth to be split into 4 if $\gamma > 19/35$. Similarly, for $\gamma<1/35$, the two cliques of \ref{fig:2cliques} are merged. The fact that the proper value of $\gamma$ may vary from cluster to cluster can render the global optimization process locally unsuccessful.
\\
\textbf{Method of Nepusz et al. \cite{Nepusz}} -- the main goal of the work is to provide community detection framework using fuzzy (soft) memberships, in order to handle overlaps. The proposed realization of the framework, when restricted to conventional hard memberships (and unweighted networks), is equivalent to the previous method with $\gamma=1$, and with a different heuristic.
\\
\textbf{Stochastic blockmodel of Hofman and Wiggins \cite{HW}} -- based on the assumption that the community structure can be fitted by a blockmodel in which intra- and inter-cluster nodes are connected with probabilities $\vartheta_c$ and $\vartheta_d$ respectively, \cite{HW} aims to minimize the Hamiltonian
\begin{equation} \label{eq:HW}
 H=-\sum_{i<j}(J_LA_{ij}-J_G)\delta_{\sigma_i, \sigma_j}-\sum_{\mu}n_{\mu}\ln\frac{n_{\mu}}{n}
\end{equation}
where $\sigma_i$ is the cluster of node $i$, $n_{\mu}$ is the size of cluster $\mu$, $J_G=\ln(1-\vartheta_d)/(1-\vartheta_c)$, $J_L=\ln\vartheta_c/\vartheta_d + J_G$. The number and sizes of clusters, the cluster members, and the probabilities $\vartheta_c$ and $\vartheta_d$ are determined by minimization. In other words, a community structure should be found which maximizes the edge densities inside the communities (the $J_LA_{ij}-J_G$ term), with the restrictions that 1) each node belongs to exactly one community; 2) the expected intra- and inter-cluster edge densities are both constants. The method of Hastings \cite{Hastings} is a special case of this method, needing $J_G$ and $J_L$ as input, and discarding the last term in equation \ref{eq:HW}. The method of Ronhovde and Nussinov \cite{RN} is also a special case with the same restrictions, i.e. being equivalent to the Hastings method. One can see immediately that equation \ref{eq:HW} defines a global method, which is realized by the global $J_G$ and $J_L$ coupling constants. Since the discovery of the resolution limit of modularity it is known that globality leads to counterintuitive local trade-offs. The situation is not different here, a simple calculation for Fig. \ref{fig:2cliques} shows that the two cliques will be merged if $(1-\vartheta_c)/(1-\vartheta_d)>2^{-12/35}(\vartheta_d/\vartheta_c)^{1/35}$, which can be approximated by $1-\vartheta_c>0.79(1-\vartheta_d)$, assuming that $(\vartheta_d/\vartheta_c)^{(1/35)}\approx1$. As $\vartheta_c$ corresponds to the intra-cluster edge probability, it is a reasonable criterion. Furthermore, it is similarly simple to show that for Fig. \ref{fig:sw_2cliques}, splitting into four is profitable if $(1-\vartheta_d)/(1-\vartheta_c)>4^{12/35}(\vartheta_c/\vartheta_d)^{19/35}$. 
\\
\textbf{Mixture model of Newman and Leicht \cite{NewmanLeicht}} --  based on some probabilistic modeling, \cite{NewmanLeicht} proposed the following log-likelihood to be maximized:
\begin{equation} \label{eq:NL}
 \mathcal{L} = \sum_{i,r}q_{ir}\left( \ln \pi_r + \sum_j A_{ij} \ln \Theta_{rj} \right)
\end{equation}
where $q_{ir}$ is the probability that node $i$ belongs to cluster $r$, $\pi_r$ is the fraction of nodes in cluster $r$, and $\Theta_{rj}$ is the probability that a randomly chosen link originating in cluster $r$ points to node $j$. Equation \ref{eq:NL} is reminiscent of the Hamiltonian of Hofman and Wiggins, although there are important differences. Nodes can have memberships in many clusters simultaneously (with the constraint that the sum of memberships is 1 for any node). Inter-cluster edges are counted for, while missing edges are never. The coupling strength between neighboring nodes is fine-tuned for each node-cluster pair. Considering Fig. \ref{fig:sw_2cliques} and assuming hard node memberships (i.e. $q_{ir}$ is 0 or 1 for all nodes), it is easy to show that splitting into 4 is favored over putting all nodes into one cluster.
\\
\textbf{Mixture model of Wang and Lai \cite{NL++}} -- Wang and Lai improved the mixture model of Newman and Leicht, arriving to the log-probability
\begin{equation}
 \mathcal{L}=\sum_{i,r}q_{i,r}\left( \ln \pi_r + \sum_j A_{ij} \ln \rho_{rj} + \sum_j (1-A_{ij}) \ln(1-\rho_{rj}) \right)
\end{equation}
where $\rho_{rj}$ is the probability that a node in cluster $r$ has a link to node $j$. Now $\mathcal{L}$ counts also the missing edges. For a hard clustering ($q_{ir}=0$ or 1) it is easy to calculate that Fig. \ref{fig:sw_2cliques} is preferred in 4 pieces over 1. 
\\
\textbf{Likelihood modularity of Bickel and Chen \cite{BickelChen}} -- the proposition is to maximize 
\begin{equation} \label{eq:BC}
Q_{LM}= \frac{1}{2}\sum_{c,d}n_{cd}\left(\frac{O_{cd}}{n_{cd}}\log \frac{O_{cd}}{n_{cd}} + \left(1-\frac{O_{cd}}{n_{cd}}\right)\log\left(1-\frac{O_{cd}}{n_{cd}}\right)\right) 
\end{equation}
where $n_{cd}=n_cn_d$ if $c \neq d$, $n_{cc}=n_c(n_c-1)$, $n_c$ is the size of cluster $c$, and $O_{cd}=\sum_{i\in c, j \in d}A_{ij}$. The expression is maximal if the clusters are cliques ($O_{cc}/n_{cc}=1$) which are totally separated ($O_{cd}/n_{cd}=0$). As $Q_{LM}$ is symmetric with respect to $O_{cd}/n_{cd}$ and $1-O_{cd}/n_{cd}$, bipartite structures can also get high scores, but here the analysis is restricted to the cluster-based optimum. First, it should be noted that $Q_{LM}$ penalizes clusters in which the edge density deviates significantly from its maximal value. Then, it is easy to calculate that it worth to cut Fig. \ref{fig:sw_2cliques} into 4 clusters. 
\\
\textbf{Stochastic blockmodel of Karrer and Newman \cite{Karrer}} -- it is similar to the previous case. The main difference in the function to be maximized (compared to equation \ref{eq:BC}) is the absence of the second logarithmic term representing the missing links, and the application of sums of degrees instead of cluster sizes. Similarly, a simple calculation shows that Fig. \ref{fig:sw_2cliques} gets higher score when split into four.

\subsection*{Other single-scale methods}

\begin{figure}[!h]
\begin{center}
\subfloat[]{\label{fig:2cliques_moduland}\includegraphics*[width=2cm]{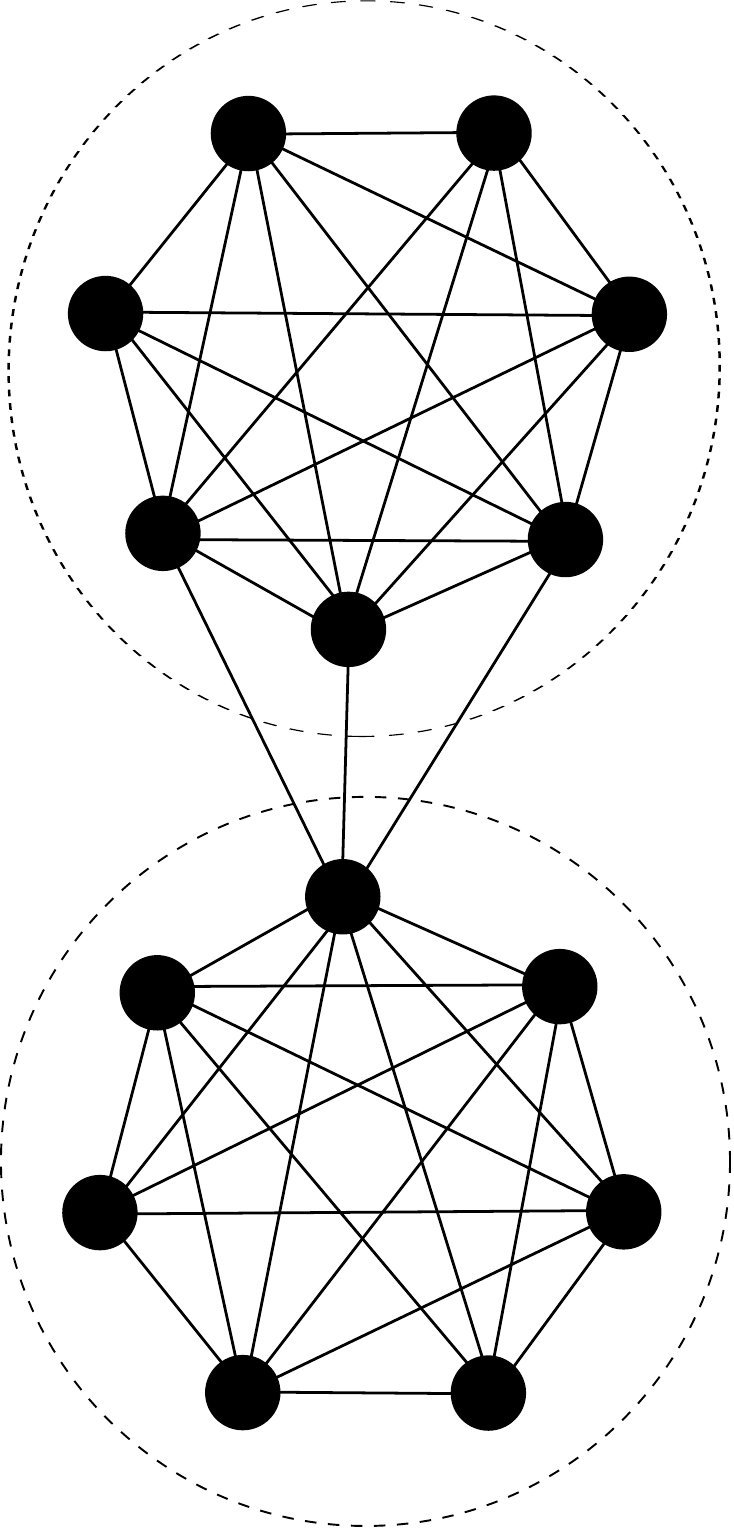}}
\qquad\qquad
\subfloat[]{\label{fig:sw_moduland}\includegraphics*[width=3.8cm]{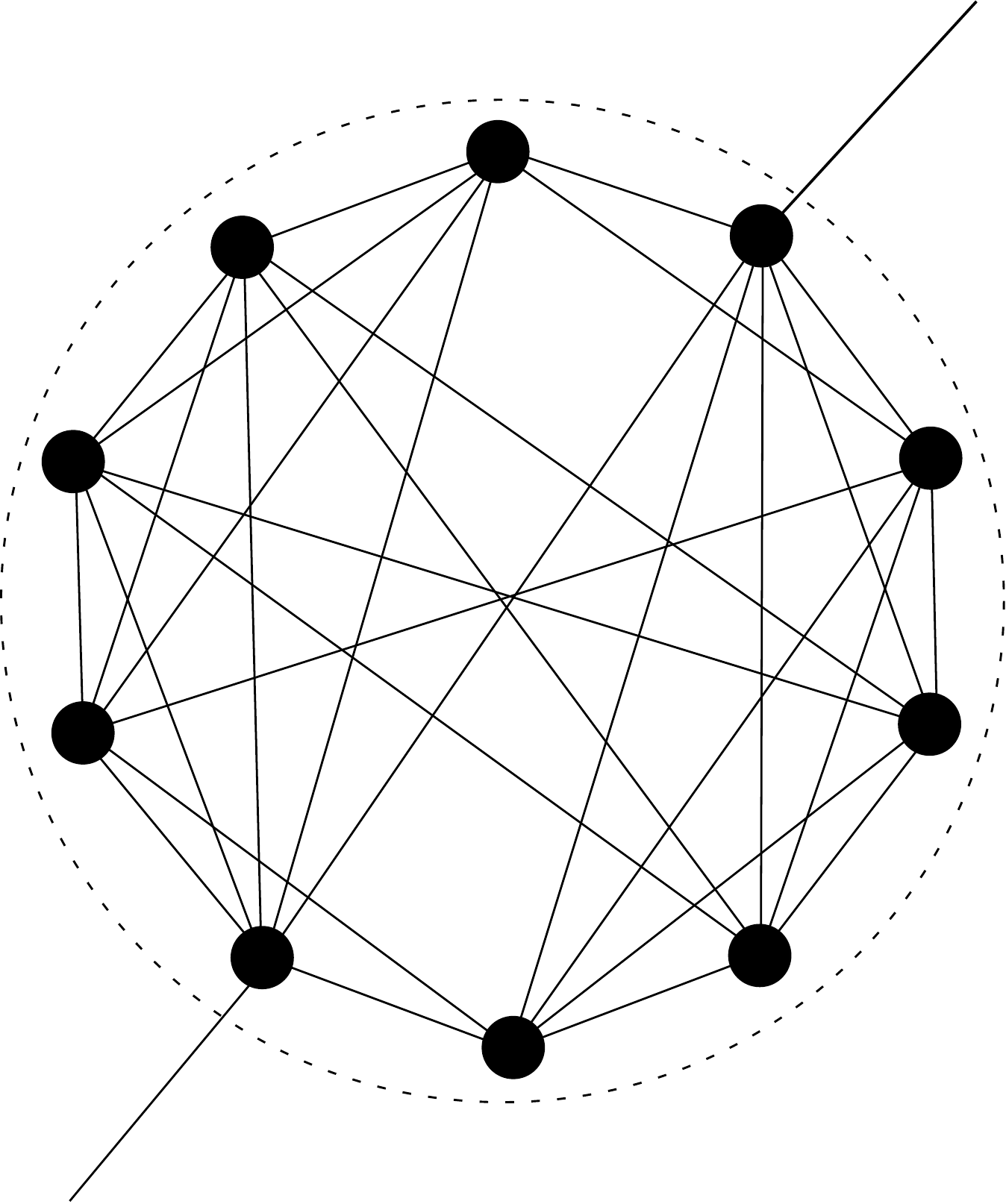}}\\
\end{center}
\caption{Subgraphs on which ModuLand-NodeLand gives counterintuitive results. Dashed lines show the desired communities.}
\label{fig:counterexample_moduland}
\end{figure}

\textbf{Infomod \cite{Infomod}} -- the aim is to compress the description of the graph, while retaining as much information as possible. The description length is given by 
\begin{equation}
 L=n\log_2 m + \frac{m(m+1)}{2}\log_2 l + \log_2 \prod_{i=1}^m\binom{n_i(n_i-1)/2}{l_{ii}}\prod_{i<j}\binom{n_i n_j}{l_{ij}}
\end{equation}
where $n$ is the number of nodes, $l$ is the number of edges, $m$ is the number of clusters. As can be seen, it is global method, where trade-offs for a global improvement may spoil local structures. And indeed, a straightforward calculation shows that for all but very small graphs  Fig. \ref{fig:2cliques} is preferred as a single community (e.g. $l \geq 128$ and $m \geq 7$). 
\\
\textbf{Method of Chauhan et al. \cite{lambda1}} -- the idea is interesting, i. e. to maximize the sum of logarithms of the largest eigenvalues of the adjacency matrices of the individual communities. However, the behavior of the largest eigenvalue of the adjacency matrix is poorly understood. As a counterexample, given a clique of size $n$, its largest eigenvalue is $n-1$, while when it is cut into two, the product of the first eigenvalues of the two $n/2-1$-sized cliques is $(n/2-1)^2$, which is larger than $n$ if $n \geq 8$ -- so it worth to cut a clique into pieces. This is the consequence of using a concave function ($\log$) in the summation, so it can be easily fixed. However, for Fig. \ref{fig:sw_2cliques}, the largest eigenvalue is $5.2$, while the largest eigenvalue of a 3-clique is 2. Summing the largest eigenvalues for the two cases (instead of summing their logarithms) results in $5.2<8$, so it worth to split Fig. \ref{fig:sw_2cliques} into 4.
\\
\textbf{Link partitioning method of Ahn et al. \cite{Ahn}} -- although the described method applies a hierarchical clustering, using an objective function (edge density of clusters) results in a single set of communities. The objective function averages the densities of all clusters, consequently it is a global quantity; its maximum does not guarantee that each cluster is optimal, just the average -- nothing prevents the over- or underpartitioning of individual clusters at the global optimum. 
\\
\textbf{Community landscape method of Kovács et al. (ModuLand) \cite{Moduland}} -- see also at the hierarchical methods. The interesting idea is to give a scalar value to the edges indicating how strongly an edge belongs to communities, then identifying the local maxima \& their surroundings (``hills``) as the communities. The scalar value for the edges is obtained as the number of appearances of the edges in some auxiliary clusters. From each node (or edge), an auxiliary cluster is grown until its fitness value cannot be increased. Fitness is chosen as simply the average in-degree of the nodes in the growing cluster. After all auxiliary communities are determined this way, each edge is assigned a value equaling the number of times it occurred in the found communities. Edges with the locally highest score are defined as community cores. Membership values are assigned to remaining edges, based on how strongly are they related to the nearby cores. The method, actually a framework for several possible methods, depends heavily on the applied fitness function of the auxiliary clusters. Here the NodeLand auxiliary clustering will be investigated. It is quite easy to engineer graphs in the spirit of Fig. \ref{fig:counterexample} which are misclustered. E.g instead of Fig. \ref{fig:2cliques}, took two 7-cliques, delete 1 link from each, and connect one node with 3 nodes from the other clique, as on Fig \ref{fig:2cliques_moduland}. The fitness of one almost-clique is $40/7$, is just below the contribution of the node in the other clique ($6/1$), so the 3 links between the cliques will be included in the community. To be precise, starting a cluster from each node, one almost-clique + the connector node will appear as a cluster $7+1/3$ times, and the other almost-clique $6+2/3$ times. Fractions correspond to different possibilities when starting from the connector node. In practice, this means that with probability $2/3$, all links will have uniform scalar values (perfectly flat landscape, i.e. a single hilltop), and with probability $1/3$, a step-like landscape (still identified as a single cluster by the method). Symmetrization to $6+1/3+2/3$ and $6+2/3+1/3$ is straightforward, by creating a second bridge node also with 3 links. Similarly, Fig. \ref{fig:sw_2cliques} can be substituted by Fig. \ref{fig:sw_moduland}. It consists of two 5-cliques with connections such that each node has 2 links to the other clique. ModuLand-NodeLand will tend to separate the two cliques, although their union is much more well-separated from the rest of the graph.

\subsection*{Hierarchical methods}

\begin{figure}[!h]
\begin{center}
\includegraphics*[width=3cm, angle=90]{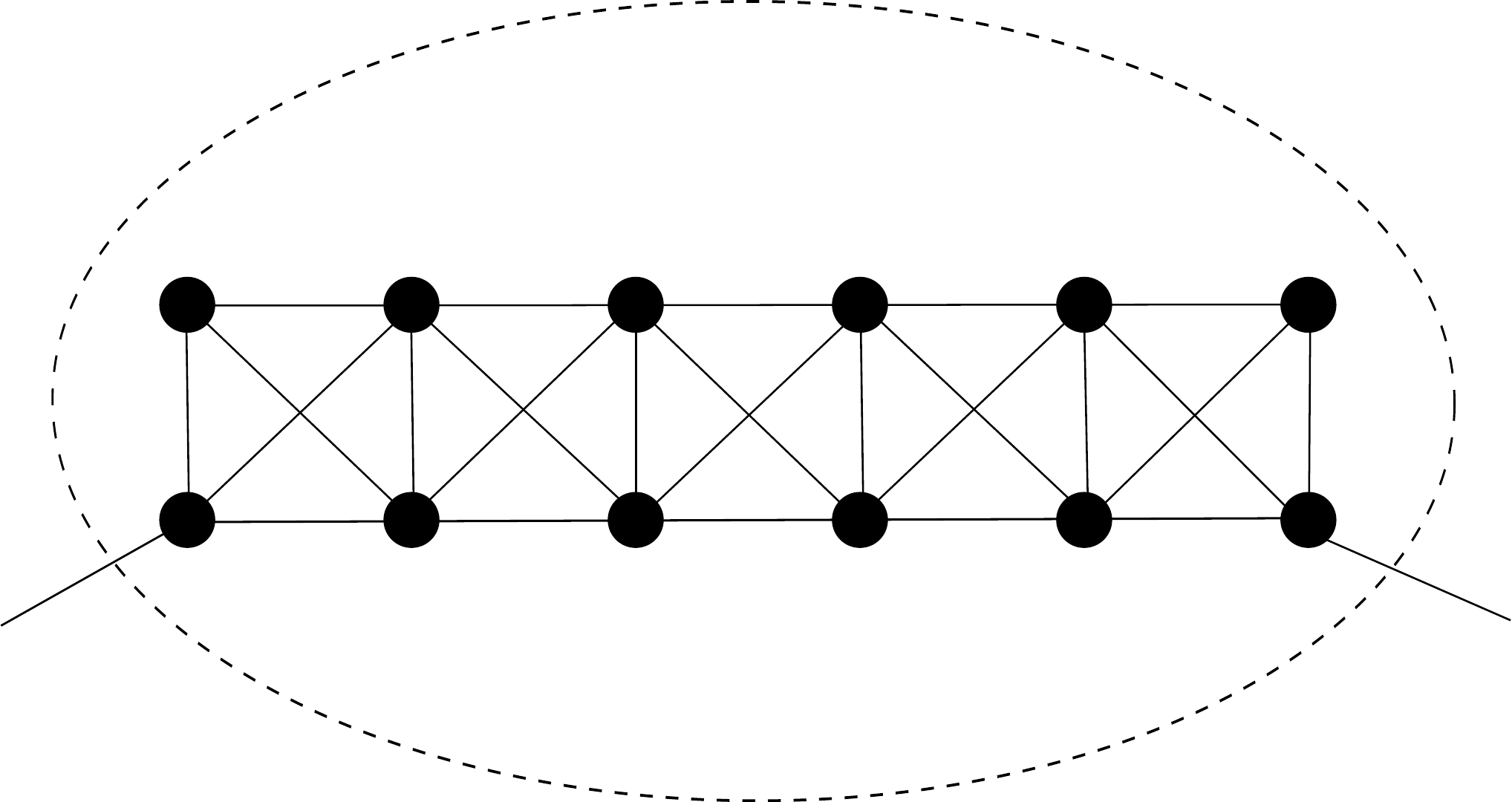}
\label{fig:chain}
\end{center}
\caption{Test case for the lowest level of the hierarchical methods.}
\label{fig:counterexample_c}
\end{figure}

Here, some hierarchical methods will be investigated. The question is whether the lowest level can be reliably used as an optimal partition (or cover). As a benchmark graph, Fig. \ref{fig:counterexample_c} will be utilized. The desired output is a single community of 12 nodes, due to their extreme separation from the rest of the graph.\\
\textbf{Method of Ruan and Zhang \cite{RuanZhang}} -- the proposition is to iteratively run modularity optimization in the found clusters, until the best modularity inside a cluster is not larger significantly than those of a corresponding random graph. Numerical calculations show that at the lowest level, Fig. \ref{fig:counterexample_c} is divided into parts\footnote{z-score is $5.3$, $Q_{\text{max}}=0.36$. Z-score is defined as the difference of the modularity of the actual graph and the modularity of a 0-model graph, divided by the variance of the modularity of the 0-model graph, z-score $=(Q-Q_{\text{0-model}})/\sigma_{\text{0-model}}$. Criterion of \cite{RuanZhang} is z-score $\geq2$, $Q_\text{{max}}\geq0.3$. Modularities were optimized using the Radatools software \cite{radatools}.}. 
\\
\textbf{Method of Sales-Pardo et al. \cite{Sales-Pardo}} -- it uses the co-occurrence of nodes in different local optima of modularity to construct a new similarity matrix, which is fitted by a block diagonal form. Communities are defined by the blocks. The method is iteratively re-applied to each community until structure deviating from a corresponding random graph is found. Again, running the method on Fig. \ref{fig:counterexample_c} results in overpartitioning (z-score of the split Fig. \ref{fig:counterexample_c} is $3.9$, the threshold used by \cite{Sales-Pardo} is $2.3$).
\\
\textbf{Hierarchical Infomap \cite{hierIM}} -- this is an extension of the Infomap method \cite{Infomap}. It is easy to calculate that, similarly to the previous cases, splitting Fig. \ref{fig:counterexample_c} on a lower hierarchical level improves the partition.
\\
\textbf{ModuLand \cite{Moduland}} -- ModuLand can also produce hierarchical structures, by iteratively re-running the clustering procedure on the network of clusters (links between clusters are defined by node overlaps). Accordingly, the lowest level clusters are the ones obtained by a simple ModuLand run, which is susceptible to mispartitioning, as described some paragraphs above.
\\
\textbf{OSLOM \cite{OSLOM}} -- the method applies statistical significance as fitness. Although its output depends to a certain degree on the whole graph, running it on Fig. \ref{fig:counterexample_c} (as the whole graph) results in a bisection. As the method tries to find the so-called minimal significant clusters,  by trying to split already found significant subgraphs while the rest of the graph is neglected, it will divide Fig. \ref{fig:counterexample_c} independently of the rest of the graph.\\

\end{document}